\documentclass[12pt]{article}
\usepackage{scicite}
\usepackage{array}
\usepackage[a4paper, text={16.5cm, 25.2cm}, centering]{geometry}
\usepackage[labelfont=bf]{caption}
\usepackage{varwidth}
\DeclareCaptionLabelSeparator{pipe}{ | }
\usepackage{amsmath}
\usepackage{booktabs,subcaption,amsfonts,dcolumn}
\usepackage{makecell}
\usepackage{url}
\usepackage{diagbox}
\usepackage{anyfontsize}
\usepackage{t1enc}
\usepackage{diagbox}
\usepackage{booktabs}
\usepackage{blindtext}
\usepackage{hyperref}
\usepackage{float}
\usepackage{tabularx}
\usepackage[dvipsnames]{xcolor}
\usepackage[linesnumbered, ruled, vlined]{algorithm2e}
\usepackage[group-separator={,},group-minimum-digits={3}]{siunitx}

\SetAlgorithmName{Supplementary Algorithm}{Supplementary Algorithm}{List of Supplementary Algorithms}

\newcolumntype{x}[1]{>{\centering\arraybackslash\hspace{0pt}}p{#1}}

\newcommand{\blue}[1]{\textcolor{black}{#1}}

\usepackage{url}
\usepackage{times}
\usepackage{gensymb}
\usepackage{lineno}

\usepackage{graphicx}
\usepackage{booktabs}
\usepackage{multirow}
\usepackage{threeparttable}
\usepackage{booktabs}    
\usepackage{enumitem}
\usepackage{bm}
\usepackage{upgreek}
\usepackage{microtype}
\usepackage[utf8]{inputenc}

\usepackage[textsize=tiny,textwidth=1.4cm]{todonotes}

\newcolumntype{M}[1]{>{\centering\arraybackslash}m{#1}}

\newcommand{\squishlist}{\begin{itemize}[itemsep=1pt,parsep=2pt,topsep=3pt,partopsep=0pt,leftmargin=0em, itemindent=1em,labelwidth=1em,labelsep=0.5em]}
\newcommand{\squishend}{\end{itemize}}

\newcommand{\header}[1]{\vskip 0.1cm \noindent{\bf #1}} 
\author{Kuang Yuan$^{1\dag}$, Freddy Yifei Liu$^{1\dag}$, Tong Xiao$^{2}$, Yiwen Song$^{1}$, Chengyi Shen$^{3}$,\\
Saksham Bhutani$^{1}$, Justin Chan$^{1\ast}$, Swarun Kumar$^{1\ast}$\\
\small
{$^{1}$Department of Electrical and Computer Engineering, Carnegie Mellon University, Pittsburgh, PA, USA}\\
\small
{$^{2}$Department of Medical Physics and Acoustics, Carl von Ossietzky Universit\"{a}t Oldenburg, Germany}\\
\small
{$^{3}$}	
College of Computer Science and Technology, Zhejiang University, Hangzhou, China\\
\small
{$^\dag$Equal contribution first authors}\\
\small
{$^\ast$Corresponding authors: justinchan@cmu.edu, swarun@cmu.edu}
}
\begin{document}

\newenvironment{sciabstract}{%
\begin{quote} \bf}
{\end{quote}}

\title{Active noise cancellation on open-ear wearable devices}
\date{}

\baselineskip24pt

\maketitle

\begin{sciabstract}
\noindent 
Active noise cancellation (ANC) is widely deployed on consumer headphones and earbuds to suppress environmental noise. However, existing ANC systems require an error microphone at the user's ear canal to measure residual sound, preventing deployment on emerging open-ear wearable devices such as smart glasses and VR headsets, which leave the ear unoccluded. \blue{Here we present an ANC system for open-ear wearables that suppresses environmental noise using only microphones and miniaturized open-ear speakers embedded within the frame of the wearables, removing the need for an in-ear error microphone. Our low-latency computational pipeline uses a neural network to estimate the noise at the ear from an array of eight microphones distributed around the wearable’s frame and generates an anti-noise signal in real-time. This mapping generalizes to unseen users and acoustic environments without prior acoustic measurement. We develop a custom glasses prototype and evaluate across eleven unseen users and eight unseen environments under mobility in the 100 to 1000~Hz frequency range, where environmental noise is concentrated.} We achieve a mean noise reduction of 9.6~dB without any calibration, and 11.2~dB with a brief user-specific calibration. Further, we demonstrate that our approach extends to the broader class of open-ear wearables including VR headsets and headbands.

\end{sciabstract}

\vskip 0.4in

\vskip 0.2in

\section*{Introduction}\label{sec:intro}

\begin{figure*}[t]
    \centering
    \includegraphics[width=0.9\linewidth]{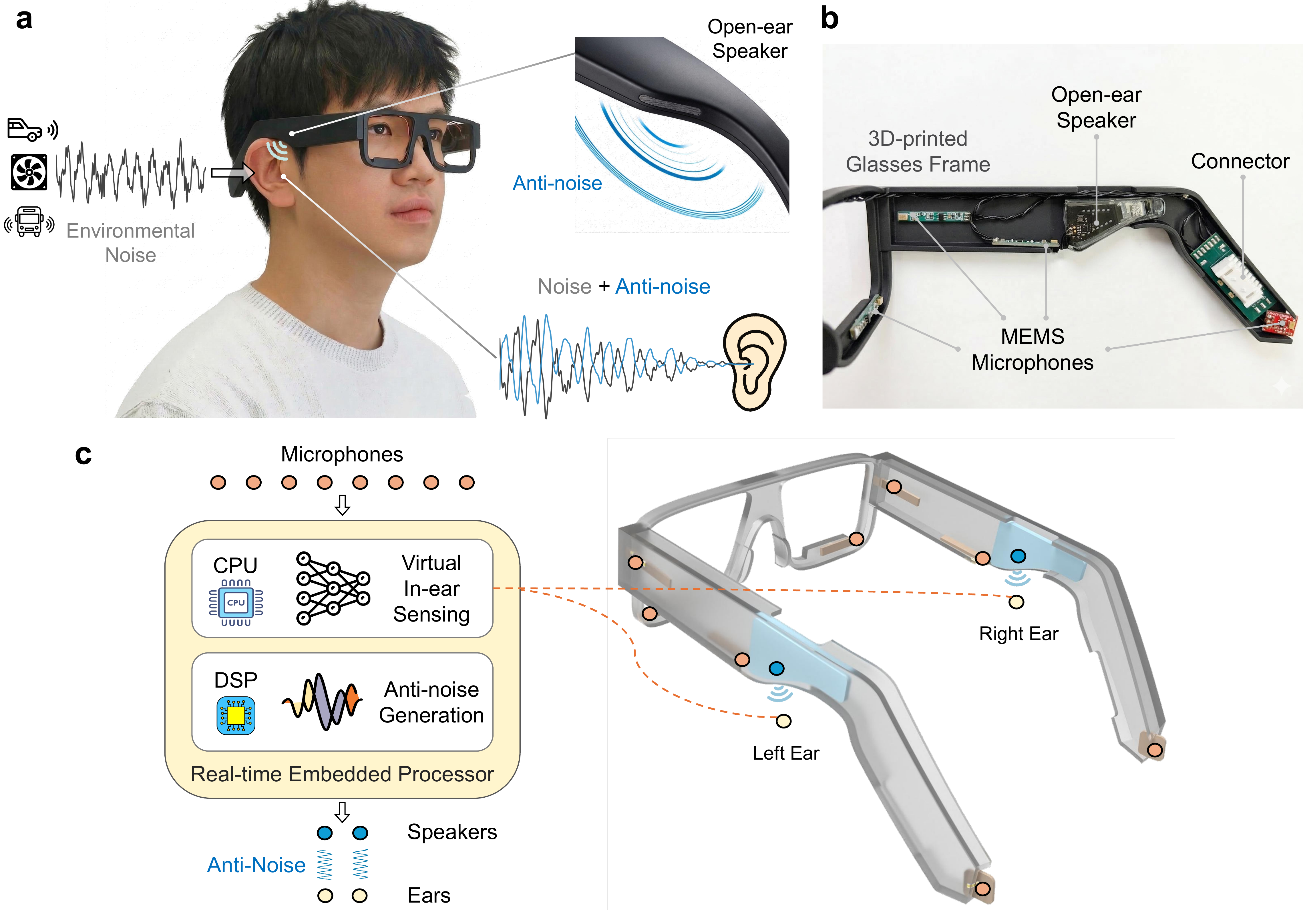}
    \caption{\textbf{Active noise cancellation for open-ear wearables} demonstrated on smart glasses. \textbf{a,} Our system suppresses environmental noise by playing anti-noise through open-ear speakers, thereby reducing noise heard at the user's ear. \textbf{b,} Interior view of glasses prototype, showing the location of microphones and an open-ear speaker mounted inside the 3D-printed glasses frame. \textbf{c,} System architecture and microphone layout. Eight microphones (orange circles) distributed across the frame capture ambient sound.  A neural network on the embedded CPU performs virtual in-ear sensing (illustrated in orange dashed lines) to estimate ANC filters for each ear; a DSP then applies these filters to generate anti-noise signals driven through the speakers (blue circles) in real time. 
    }
    \label{fig:concept}
\end{figure*}


Active noise cancellation (ANC), first proposed in the 1930s~\cite{lueg1936}, has evolved to become the dominant approach for suppressing unwanted environmental noise across aviation headsets~\cite{boseA30AviationHeadset}, telecommunication systems~\cite{EPOStelecomm}, and consumer headphones and earbuds~\cite{airpods4}. In each of these systems, effective cancellation has relied on a closed acoustic coupling to the ear, in which an \textit{error microphone} positioned inside or at the entrance of the ear canal measures residual noise and provides feedback for cancellation~\cite{nelson1991active, kuo1999active}. This requirement has confined effective ANC to ear-occluding form factors such as headphones and earbuds.

Consequently, conventional ANC is fundamentally incompatible with open-ear devices such as smart glasses and AR/VR headsets~\cite{raybanmeta, googlexrglasses, rokid} which deliberately leave the ear unoccluded to preserve the user's ambient awareness and social comfort for everyday use.
Instead, they deliver audio through open-ear speakers mounted near the temples that direct sound towards the ear~(Fig.~\ref{fig:concept}a). However, the open acoustic path does not provide physical isolation, allowing undesired environmental noise to interfere with desired audio playback at the ear, degrading perceptual audio quality, particularly in loud settings such as busy streets, caf\'es, or public transit. Achieving ANC without occluding the ear would therefore enable noise-robust auditory experiences for streaming music, phone calls, and voice-based artificial intelligence (AI) interactions.



\blue{Prior efforts toward open-ear ANC have largely followed the local active control paradigm, such as with active headrests \cite{headrest_1999, Jung2019local} and their head-tracked extensions \cite{elliott2018head, Veronesi2025interpolation, Xiao2020ultra}. These systems measure the acoustic plant responses in advance for a known fixed geometry, which restricts cancellation to a bounded quiet zone. They also rely on external infrastructure such as external head trackers. Remote microphone techniques ~\cite{shi2019selective, shi2020active, Antonanzas2023remote} similarly rely on sensors placed in the environment for a fixed zone of quiet. None of these assumptions hold for a mobile wearable operating in arbitrary, uncharacterized environments. More recently, deep-neural-network~(DNN)-based algorithms have been proposed to accelerate ANC adaptation across noise types ~\cite{shi2020feedforward, luo2025deep, wang2025transferable, luo2025frequency}. However, the neural networks in these systems only adapt the control filter to noise content. The knowledge of the acoustic path is still derived from an in-ear error microphone, and must be re-acquired in each new environment. Existing architectures cannot estimate the acoustic paths themselves without the use of an error microphone. Although open-ear wearable devices are a rapidly growing segment of personal audio devices \cite{smartglasses_growth}, no existing ANC architecture can support their unoccluded design.}


Here, we present an ANC system on open-ear smart wearables that provides effective noise reduction across various real-world acoustic environments without any error microphone at the ear (Fig.~\ref{fig:concept}a). Rather than placing dedicated sensors in the environments, we observe that modern open-ear wearables such as smart glasses have already begun integrating multiple microphones distributed around the frame for spatial audio capture and voice assistants~\cite{raybanmeta, googlexrglasses, rokid}, providing an untapped opportunity for ANC. \blue{Our main finding is that a neural network can estimate the sound at the ear from these frame microphones alone, by leveraging the approximately fixed spatial relationship imposed by the rigid glasses frame geometry. Importantly, this learned mapping generalizes to unseen users and unseen acoustic environments without any prior measurement, thereby enabling effective and practical ANC on open-ear smart wearables without an in-ear error microphone.} 

We prototype our design on a custom 3D-printed glasses frame that integrates eight miniaturized micro-electro-mechanical-systems (MEMS) microphones and two open-ear speakers~(Fig.~\ref{fig:concept}b). At the core of our system is a computational pipeline that performs \textit{virtual in-ear sensing} to predict the noise at the ear using the microphone signals from across the glasses frame as input to a neural network which estimates a set of learnable filters. The estimated filters are then applied on a dedicated digital signal processing~(DSP) unit, which generates anti-noise to reduce environmental noise at the user's ears (Fig.~\ref{fig:concept}c). The DSP operates with an end-to-end processing latency of \qty{113}{\micro\second}, while the neural network updates the filter coefficients every \qty{200}{\milli\second} to adapt to dynamic acoustic conditions in the real-world. 
Beyond smart glasses, we show how our findings generalize to the broader class of open-ear wearables, including AR/VR headsets and smart headbands, introducing the ANC capability previously absent in these devices. 
\blue{More broadly, our system lays the groundwork for next-generation programmable auditory experiences on open-ear wearable devices, such as spatially selective ANC~\cite{xiao2023spatially, xiao2024icassp, Xiao2025fa, Xiao2025soft}, semantically selective listening~\cite{veluri2023semantic}, and personalized sound zones~\cite{Chen2024soundbubble}, in which users suppress unwanted noise while remaining aware of the desired sounds around them.}

\section*{System design} \label{sec:system_design}

\begin{figure*}[!t]
    \centering
    \includegraphics[width=\linewidth]{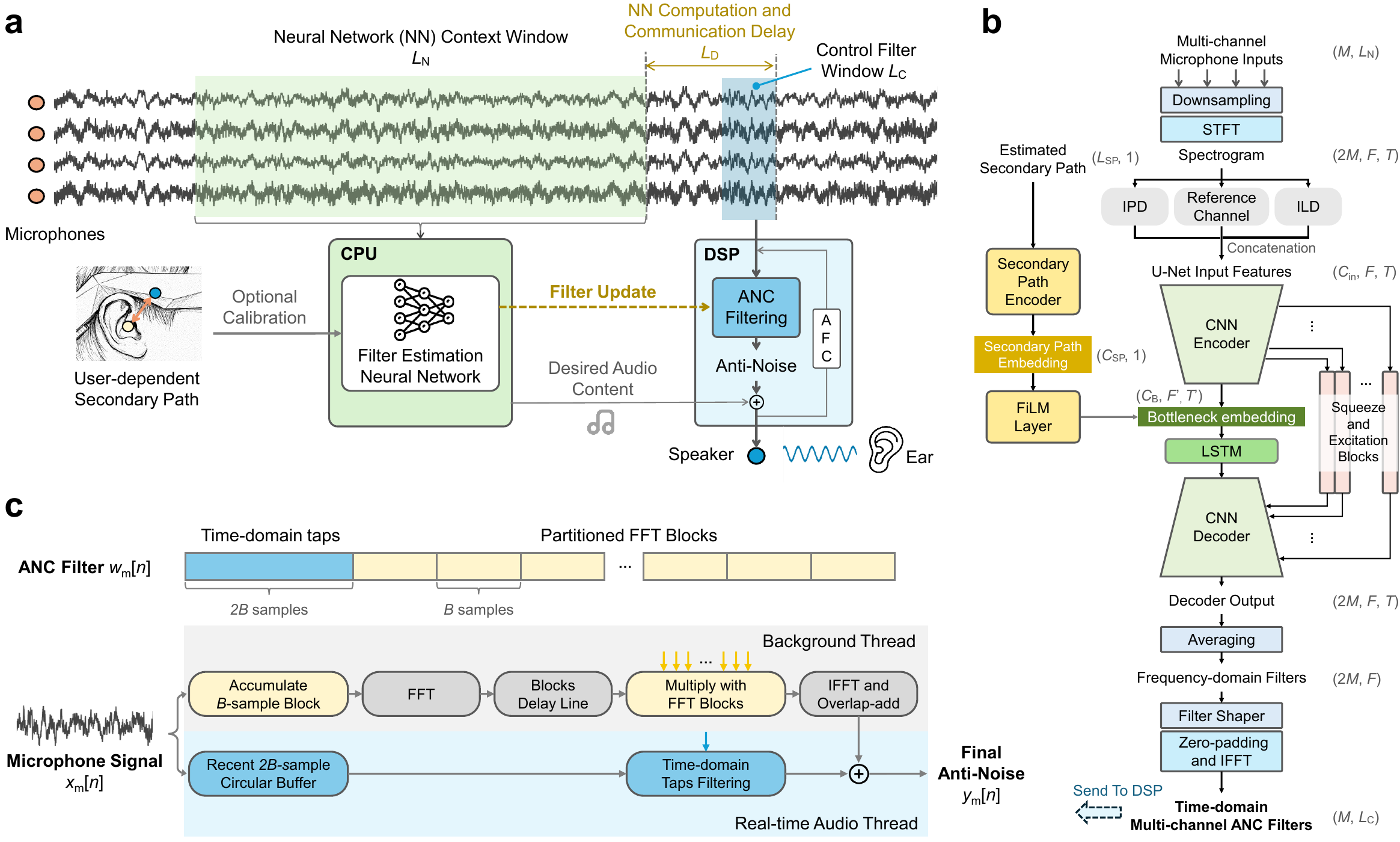}
    \caption{\textbf{System architecture. a, Overview of the dual-pipeline design} (illustrated for one side with four microphones and one open-ear speaker). Microphone signals from the frame are input to a neural network on the CPU that estimates ANC filter coefficients. These filters are passed to the DSP unit, which generates anti-noise from the most recent microphone signals and combines it with any desired audio content at the speaker.  \textbf{b, Neural network architecture.} Multi-channel microphone inputs are downsampled and transformed via STFT into spectrograms, from which IPD, ILD, and a reference channel are extracted as input features. A U-Net with a CNN encoder, squeeze-and-excitation blocks, and an LSTM at the bottleneck processes these features. A pre-calibrated secondary path estimate is encoded and injected into the bottleneck via a FiLM layer, conditioning the network on user-specific ear acoustics. The decoder outputs frequency-domain filters, which are averaged across time, shaped, and converted to time-domain multi-channel ANC filters via IFFT before being sent to the DSP. \blue{\textbf{c, Hybrid partitioned convolution on the DSP.} The ANC filter is split into a direct segment applied in the time domain for low latency, and longer partitioned segments applied via frequency-domain multiplication for better efficiency. These two results are summed to produce the final anti-noise signal.}}
    \label{fig:architecture}
\end{figure*}

{Achieving ANC on open-ear wearables requires solving two coupled problems that operate on different timescales.
\textit{First,} the system must estimate how sound propagates from the environment to the user's ear canal, a mapping that depends on the noise characteristics, source geometry, head diffraction, and individual ear anatomy.
This mapping evolves continuously as the user moves or acoustic conditions change, but varies gradually over hundreds of milliseconds.}
\textit{Second,} the system must simultaneously apply these propagation estimates to generate anti-noise waveforms with sub-millisecond latency, because the acoustic propagation delay between the glasses frame microphones and the ear is typically less than a few hundred microseconds, and any additional processing delay introduces phase errors that degrade cancellation~\cite{kuo1999active, elliott1993active}.
Our system addresses this through a dual-pipeline architecture that separates the two operations onto two parallel units on the processor (Fig.~\ref{fig:architecture}a).

\subsection*{Neural network-based virtual in-ear sensing}

As noise propagates toward the user, it diffracts around the head, creating correlated acoustic measurements at both the frame microphones and the ear canal. Because glasses maintain a fixed geometric relationship with the user's head, these correlations can be learned from the spatial cues provided by the eight microphones distributed across the frame. We model this relationship as a set of relative transfer functions, represented as finite impulse response (FIR) ANC filters in the time domain that map each frame microphone signal to the signal at the ear canal, a formulation widely adopted in ANC systems to approximate acoustic propagation as a locally linear, time-varying system~\cite{kuo1999active}. 

{A neural network running on an embedded CPU (Raspberry Pi 5) estimates the FIR filter coefficients (Fig.~\ref{fig:architecture}a).} It takes $L_N =$~\qty{2}{\second} of audio from the multi-channel frame microphones as context and estimates $L_C = 2048$-tap ANC filter coefficients for each microphone-to-ear pair.
The eight microphones and two speakers are divided into two independent groups for each ear, with four microphones and one speaker per side, yielding four filters per ear.
{Given the overall maximum measured latency to compute the filters and communicate them to the DSP of \qty{161}{\milli\second} (Fig.~\ref{fig:roomsruntime}b), the neural network updates the filter coefficients every $L_D =$~\qty{200}{\milli\second}.}

{Our network (Fig.~\ref{fig:architecture}b) first downsamples the microphone signals from \qty{22050}{\hertz} to \qty{8820}{\hertz}, since active cancellation operates primarily in the low-frequency range, reducing the computational cost of subsequent neural network processing.} The downsampled signals are transformed via the short-time Fourier transform (STFT), from which three complementary input features are extracted: the interchannel phase difference (IPD), the interchannel level difference (ILD)~\cite{blauert1997spatial}, and the spectrogram of a chosen reference microphone channel. Because the acoustic transfer functions depend on the spatial relationship between noise sources and the user's head, IPD and ILD provide the network with geometric information necessary for accurate estimation, while the reference spectrogram captures the spectral content of the noise. We use the microphone closest to the open-ear speaker as the reference channel and compute IPD and ILD of all other channels with respect to it.

These features are passed to a convolutional encoder--decoder architecture with U-Net skip connections, incorporating squeeze-and-excitation blocks for channel-wise recalibration and a long short-term memory (LSTM) layer at the bottleneck to maintain temporal coherence across successive estimation windows. \blue{The decoder predicts ANC filters in the frequency domain, where acoustic responses exhibit smoother structure than in the time domain~\cite{lan2024acoustic}. For a given context window, the decoder produces one filter estimate per STFT time frame. These per-frame estimates are pooled by taking the mean along the time axis to improve estimation stability, yielding a single filter per channel per forward pass.} The resulting frequency-domain filters are then passed through a half-cosine roll-off shaper to suppress edge artifacts, zero-padded to restore the original \qty{22050}{\hertz} sample rate, and converted to time-domain FIR filters via the inverse fast Fourier transform (IFFT) before being sent to the DSP.


{The predicted filters must also account for the \textit{secondary path}, the acoustic signal path between the open-ear speaker and the user's ear canal, which varies across individuals due to differences in head geometry and ear shape. Our system supports two modes: a population-averaged secondary path estimate that requires no calibration, or a user-specific estimate obtained through a brief calibration with a temporary in-ear microphone, without model retraining. During inference, this estimate is compressed by a learned encoder into a fixed-dimensional embedding and injected into the network bottleneck via a feature-wise linear modulation (FiLM) layer to condition the decoder.}

\subsection*{Real-time anti-noise generation}

{Once the neural network has estimated the ANC filter coefficients, a dedicated ultra-low-latency DSP unit (Bela) executes them in real time to generate the anti-noise signal (Fig.~\ref{fig:architecture}c). As the DSP produces the anti-noise waveform sample-by-sample, our sampling rate of \qty{22050}{\hertz} imposes a per-sample computation budget of \qty{45}{\micro\second}. To meet this constraint, we employ a hybrid partitioned convolution scheme~\cite{wefers2015partitioned} that achieves a median processing latency of \qty{38}{\micro\second} for our chosen filter length of  $L_C=2048$-taps (Fig.~\ref{fig:roomsruntime}c), with the per-sample budget of \qty{45}{\micro\second} serving as the hard upper bound on computation latency.}

\blue{This hybrid partitioning scheme splits each 2048-tap ANC filter into two segments processed on parallel threads. The first $2B$ ($B = 128$ samples) taps, the \textit{direct segment},  are applied as a direct time-domain convolution on the real-time audio thread, operating on a circular buffer of the most recent $2B$ input samples. The remaining taps, the \textit{partitioned segments}, are blocks of $B$ samples and processed on a background thread via partitioned frequency-domain convolution. Each new $B$-sample block is transformed via FFT and appended to a frequency-domain delay line containing recent input blocks. The transformed blocks in this delay line are then multiplied by their corresponding pre-transformed filter partitions, summed in the frequency domain, and reconstructed via IFFT with overlap-add. The final anti-noise signal is produced sample-by-sample by adding the direct segment output to the appropriate sample from the latest available partitioned segment.}

The DSP also combines the anti-noise with any desired audio playback content, such as speech or music, before driving the open-ear speaker. To prevent the anti-noise emitted by the speakers from contaminating the reference microphone signals, the DSP incorporates an acoustic feedback cancellation~(AFC) stage that subtracts the predicted speaker-to-microphone coupling from the raw microphone input before ANC processing.

{We empirically characterized the median end-to-end latency from microphone input to speaker output to be \qty{113}{\micro\second}, with \qty{45}{\micro\second} as the fixed computation time and the remaining \qty{68}{\micro\second} to be the overhead associated with other elements of the audio chain including the analog-to-digital converter (ADC) and digital-to-analog converter (DAC).}

\begin{figure*}[!t]
    \centering
    \includegraphics[width=1\linewidth]{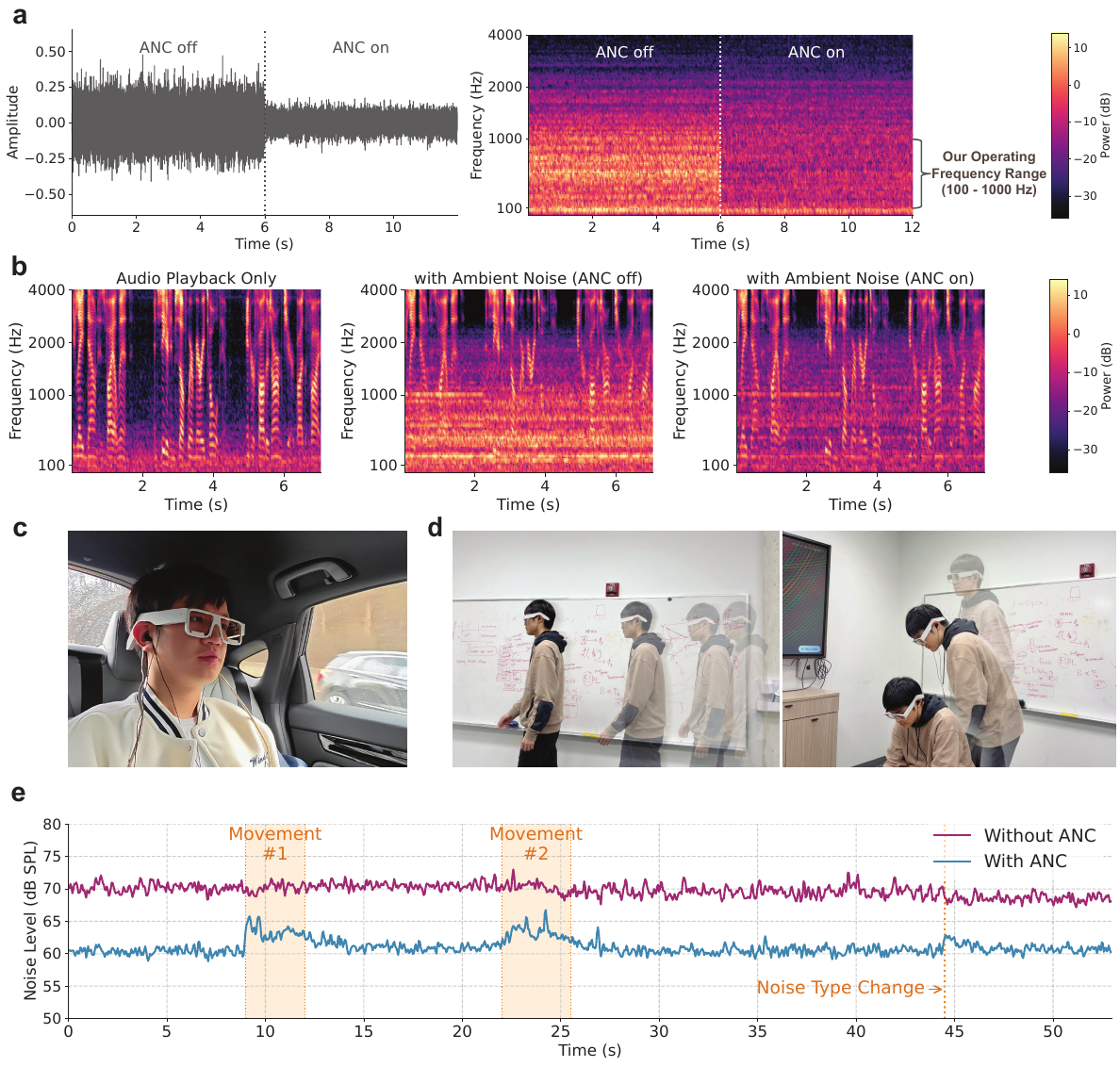}
    \caption{\textbf{ANC system working in real-world. a,} ANC reduces broadband noise at the user's ear, with the most pronounced attenuation in the 100--1000~Hz operating range as seen in the spectrogram. \textbf{b,} ANC suppresses ambient noise during simultaneous speech playback, and enhances the spectral clarity of speech content. \textbf{c,} Demonstration in a real-world car environment. The user wears our glasses prototype for real-time noise reduction, and in-ear mics for ANC performance monitoring. \textbf{d,} Demonstration during user movement. \textbf{e,} Noise level at the ear over time, with and without ANC. ANC maintains $\sim$10~dB reduction across user movements and a mid-recording noise type change.}
    \label{fig:ANC_example}
\end{figure*}

\section*{End-to-end demonstration}

\begin{figure*}[!htbp]
    \centering
    \includegraphics[width=0.9\linewidth]{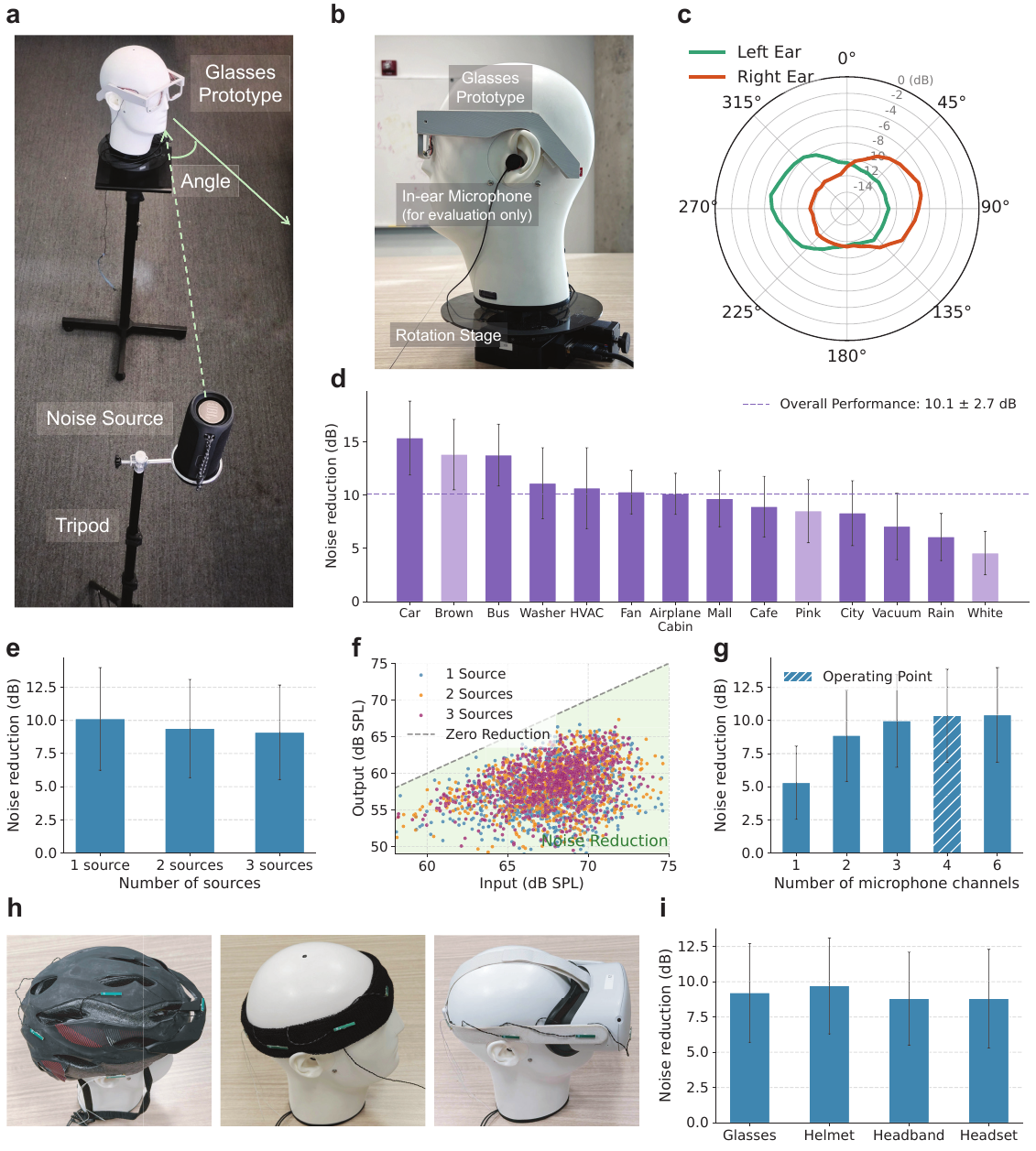}
    \caption{\blue{\textbf{Evaluation on data collected from mannequin head.}} \textbf{a,} Automated data collection setup using a mannequin head with a height adjustable noise source arriving from different directions. \textbf{b,} Close-up of mannequin head wearing our glasses prototype and an in-ear microphone, placed on a rotating stage. \textbf{c,} Noise reduction across directions-of-arrival for both left and right ears. \blue{A smaller radius indicates greater noise reduction. \textbf{d,} Mean noise power reduction across different noise types. Real world noises are shown in dark purple, whereas colored noises are shown in light purple. The error bar is the standard deviation. \textbf{e,} Noise reduction over different number of noise sources in the environment. \textbf{f,} Noise reduction for noises of different levels. Performance is shown for 1, 2, and 3 simultaneous noise sources in the environment. Points below the dashed line indicate noise reduction. No evaluation segment falls above the zero-reduction line, meaning the system never produces net noise enhancement in any tested condition.} \textbf{g,} Effect of the number of reference microphones used on system performance (single ear). The hatched bar represents the configuration we use in practice. \blue{\textbf{h,} Open-ear wearable prototypes that span a range of microphone geometries, including a smart helmet, a smart headband, and a VR headset. \textbf{i,} Noise reduction performance of our framework applied across these geometries (glasses for reference), evaluated in simulation. Synthetic data was only used for this study. The framework was additionally validated on physical prototypes and real-world noise (see text).}}
    \label{fig:mannequin}
\end{figure*}

To demonstrate the end-to-end viability of our system, we implemented the complete pipeline primarily on a custom glasses prototype (see Fig.~\ref{fig:mannequin} for generalizations to other kinds of wearables). In practice, the system effectively attenuates broadband ambient noise in the 100–1000 Hz operating range across real-world scenarios (Fig.~\ref{fig:ANC_example}a). When the open-ear speaker simultaneously delivers desired audio content such as speech or music in a noisy environment (Fig.~\ref{fig:ANC_example}c), the system suppresses the ambient noise and enhances the spectral clarity of the playback signal (Fig.~\ref{fig:ANC_example}b and Supplementary Videos 1, 2). The neural network updates ANC filters every \qty{200}{\milli\second}, allowing the system to continuously adapt to changes in the user's position, head orientation, noise source, and surrounding acoustic environment. As shown in Fig.~\ref{fig:ANC_example}d–e and Supplementary Video 3, the system maintains approximately 10 dB of noise reduction in a car environment and during deliberate user movements, recovering quickly after each movement and sustaining consistent reduction even when the noise source type changes mid-recording.

\section*{Evaluation} 
\label{sec:evaluation}

We evaluated our open-ear ANC system in three stages, primarily conducted on a pair of prototyped smart glasses, accompanied with results showcasing generality to other types of wearables. First, we performed controlled benchtop experiments on an acoustic mannequin head to characterize the system under well-defined conditions, including different noise directions of arrival (DOAs), noise types, microphone configurations, and numbers of noise sources. Next, we evaluated the system in real-world settings across 11 unseen users and 8 unseen environments to assess generalization to variability in head geometry, glasses fit, and environmental conditions. Finally, we examined how the system affects the desired audio playback by measuring speech and music enhancement using both objective metrics and subjective user ratings of clarity and noise reduction.

\noindent \textit{Selection of noise sources.} We evaluated our system across 11 noise types, primarily consisting of environmental noises commonly encountered in daily life, including transportation (car, bus, airplane cabin), indoor appliances (washer, vacuum, fan, HVAC), and ambient soundscapes (café, mall, city, rain). We define our target bandwidth as 100–1000 Hz, which captures the dominant energy of most everyday environmental noise~\cite{berglund1999guidelines, salomons2011practical} and aligns with the effective operating range of prior ANC systems~\cite{Xiao2020ultra, elliott1993active}.

\subsection*{Benchtop evaluation on mannequin head}

Effective training of the proposed system requires a large dataset spanning diverse spatial and noise source configurations to ensure robust generalization. We first collected a large-scale dataset using an acoustic mannequin head across five environments with varied noise source positions, heights, and azimuth angles swept by a stepper-motor-driven rotating stage (Fig.~\ref{fig:mannequin}a,b). For each acoustic scene, we simultaneously recorded signals at the in-ear microphone as the ground truth and at the frame microphones, and trained the network to map the frame microphone signals to the in-ear microphone signals. This dataset was used to pre-train the neural network (see \nameref{sec:methods}). 

This mannequin-based setup also established a controlled testbed for characterizing system performance. We reserved a held-out subset from two environments that were not seen during training for the following evaluation.


\noindent \textit{Effect of directions of arrival.} We evaluated noise reduction performance as a function of noise source DOA, with results for the left and right ear shown in Fig.~\ref{fig:mannequin}c. Across all angles, the system achieves a mean reduction greater than 6~dB. Performance varies with the propagation delay from the frame microphones to the ear. For noise sources on the same side as the ear being evaluated, the noise arrives at the ear sooner relative to the frame microphones, leaving less time for the system to generate the anti-noise. Consistent with this geometry, the left ear shows its lowest reduction near $270^\circ$, whereas the right ear shows its lowest reduction near $90^\circ$. At other angles, the longer source-to-ear propagation delays provide more lookahead and enable stronger reduction.

\noindent \textit{Performance across noise types.} Our system achieved noise reduction across all tested noise categories, with a mean reduction of $10.1 \pm 2.7$~dB across the 11 real-world noise types (Fig.~\ref{fig:mannequin}d). Performance varied with the spectral characteristics of each noise type: transportation noises with dominant low-frequency content achieved the highest reduction (car: 15.3~dB; bus: 13.6~dB), while noises with more distributed spectral energy showed moderate reduction (airplane cabin: 10.1~dB; café: 8.9~dB), and noises with substantial high-frequency content were more challenging (vacuum: 7.0~dB; rain: 6.0~dB). This frequency-dependent performance is illustrated in Extended Data Fig.~\ref{fig:noise_type_psd}, where noises dominated by low-frequency energy exhibit a larger gap between the ANC-off and ANC-on conditions, consistent with the principle that ANC is more effective at low frequencies where longer wavelengths reduce sensitivity to phase errors~\cite{kuo1999active, elliott1993active}. \blue{We additionally evaluated our system on colored noise with known spectra under the same conditions. Consistent with what we saw with real-world noise, reduction fell as spectral energy shifted towards the higher frequencies, from 13.8 dB for brown noise to 8.5 dB for pink noise and 4.6 dB for white noise. White noise represents a worst case, as its flat spectrum places the most energy at high frequencies and its samples are temporally uncorrelated, meaning any gap in acoustic lookahead cannot be made up by waveform prediction.}

\noindent \textit{Effect of number of noise sources.} \blue{Fig.~\ref{fig:mannequin}e shows noise reduction performance with 1, 2, and 3 simultaneous noise sources. Our system maintains mean noise reduction above 9.0~dB in all three cases, indicating that the system is robust to the increased acoustic complexity introduced by multiple concurrent noise sources. Fig. \ref{fig:mannequin}f shows the noise reduction performance on the individual evaluation segments with 1, 2, and 3 simultaneous noise sources across a range of input sound levels. All measurements fall below the zero-reduction line, meaning the system never produces net noise enhancement. This is an important safety property for any ANC system operating without an error microphone, and confirms that the predicted filters do not amplify noise in any tested condition. Output levels ranged approximately 50--65~dB~SPL for inputs spanning 60--75~dB~SPL, showing reduction across a wide range of conditions. }

\noindent \textit{Effect of number of microphones.} We evaluated the noise reduction performance for a single ear (averaged from both sides) as a function of the number of microphones on the corresponding side of the frame. For each channel count, we selected the best-performing channel combination out of all possibilities (Extended Data Fig.~\ref{fig:mic_config}a). Fig.~\ref{fig:mannequin}g shows that performance improves with increasing microphone count: mean reduction increased from 5.3~dB with a single microphone to 8.9, 10.0, and 10.4~dB with two, three, and four microphones respectively, plateauing at 10.4~dB with six. This improvement arises because additional microphones provide richer spatial information about the sound field around the head. Notably, microphone placement also affects directional performance—configurations with greater spatial spread yield more uniform reduction across DOAs, as microphones closer to the noise source provide earlier acoustic look-ahead for anti-noise generation (Extended Data Fig.~\ref{fig:mic_config}b,c). However, each additional microphone linearly increases the computational load on the DSP. We therefore select four microphones per side (highlighted by the hatched bar) as our operating point, balancing performance against computational cost, and use this configuration for all subsequent evaluations.

\noindent \blue{\textit{Oracle Wiener filter benchmark.} We benchmark our system against a conventional multichannel Wiener filter solution (see Methods). This filter is computed using the ground-truth in-ear microphone, and therefore it is not a deployable baseline but instead an oracle solution that upper-bounds the reduction attainable by our system. The Wiener filter achieves a mean noise reduction performance of 14.7 dB across all noises (Extended Data Table \ref{tab:wiener}). Our network is therefore able to operate within 4.3 dB of the oracle without the use of any in-ear microphone.}

\noindent \blue{\textit{Effect of source loudspeaker hardware.} All noise sources in our main evaluation above were played through a JBL Flip 5 bluetooth speaker. To demonstrate that the specific hardware of the noise source has little impact on the noise reduction performance of our system, we conducted an additional experiment to test the performance of our system with different speakers playing the ambient noise. We tested the Genelec 8010, JBL Flip 5, and Music Legend T18 speakers. Extended Data Fig. \ref{fig:source_speaker}a shows the different frequency responses of the three speakers, and Extended Data Fig. \ref{fig:source_speaker}b shows that our noise reduction performance is consistent regardless of which of the three speakers is used, with the noise reduction across the three speakers falling within 1.2 dB of each other.}

\noindent \blue{\textit{Performance in anechoic conditions.} We further evaluated the mannequin setup in an anechoic chamber across the full azimuth range. The system achieved a mean noise reduction of 9.1 dB, compared to 10.1 dB average in real-world reverberant rooms. We attribute this difference to the composition of the incident sound field. In a reverberant room, a large fraction of energy arriving at the ear consists of reflections that arrive later than the direct-path component, providing the system with additional acoustic lookahead. In an anechoic chamber, essentially all incident energy arrives from the direct path, which is the most latency-constrained component of the field. We therefore believe this result represents a conservative lower bound on performance with respect to room acoustics.}

\noindent \blue{\textit{Generalization across microphone geometries.} Our system leverages the fixed spatial relationship between the frame microphones and the ear to perform virtual in-ear sensing. Altering the microphone layout changes the source-to-reference lookahead, secondary and feedback paths, and spatial cues (IPD, ILD) available to our model. We demonstrate that our approach extends across different microphone geometries by testing on different open-ear wearable form factors, including VR headsets, smart headbands and smart helmets (Fig.~\ref{fig:mannequin}h). Using varying microphone layouts, we evaluated the performance on synthetic data. We simulated 11,000 rooms with random dimensions, reverberation times, and up to 3 noise sources. Fig.~\ref{fig:mannequin}i shows that across all simulated devices, our mean noise reduction remains above 8.8~dB. We emphasize that synthetic data was used for this generalization study only. All other results reported elsewhere in this paper are trained and evaluated entirely on real recordings.}

We developed real-world prototypes for each of the form factors (Fig.~\ref{fig:mannequin}h) and evaluated each of their performances on a smaller set of real-world data. Our systems achieve a 11.3~dB, 10.9~dB, and 10.8~dB mean noise reduction on the helmet, VR headset and headband form factors respectively. This shows that our approach successfully generalizes across different open-ear form factors and microphone configurations.

\subsection*{Real-world evaluation across users}
\begin{figure*}[!ht]
    \centering
    \includegraphics[width=1\linewidth]{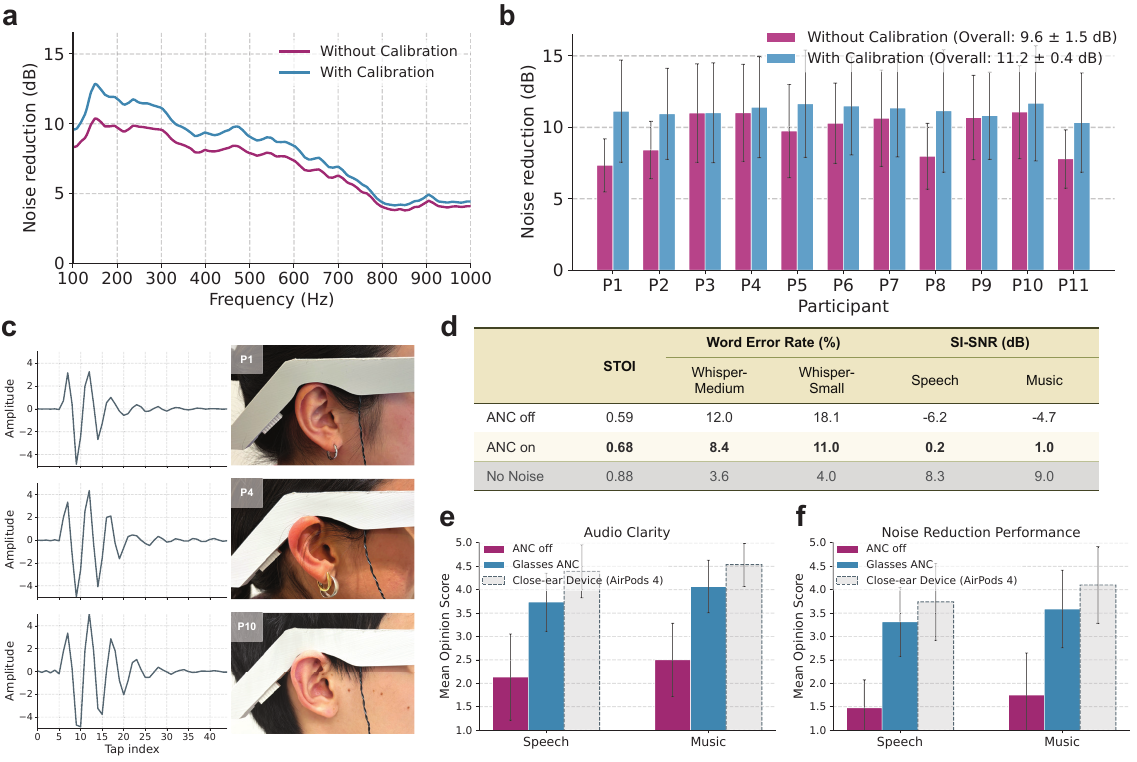}
    \caption{\textbf{Real-world evaluation across users.} We evaluate the performance of our system on 11 unseen users. \textbf{a,} Mean noise reduction performance across frequency, averaged over all users. We show performance with and without user-specific secondary path calibration. \textbf{b,} Mean noise reduction for each participant, with and without user-specific calibration. The error bars show the standard deviation. \textbf{c,} Examples of estimated secondary path impulse responses across different participants, illustrating the inter-subject variability induced by anthropometric differences and glasses fit. \blue{\textbf{d,} Objective evaluation based on audio recorded at the ear during playback of speech and music through the glasses in noisy environments, with ANC on and off. The No Noise row is a clean-speech reference recorded on a mannequin head in a quiet environment (see Methods). We report short-time objective intelligibility (STOI), automatic speech recognition (ASR) word error rate, and desired-signal scale-invariant signal-to-noise ratio (SI-SNR).} \textbf{e, f,} Subjective user experience ratings of \textbf{e,} audio clarity \textbf{f,} and noise reduction of our glasses-based ANC system, compared with ANC disabled and AirPods 4. AirPods 4, as a close-ear device, is included for qualitative reference only, as its in-ear error microphone and ear-located loudspeaker make it not directly comparable to our open-ear glasses platform.}
    \label{fig:usereval}
\end{figure*}

While the mannequin experiments provided controlled validation of our framework, real-world deployment introduces additional variability including differences in head shape, ear geometry, and how the wearables sit on each wearer, which affect both the head-related transfer functions (HRTFs) that govern sound propagation around the head and the secondary path between the open-ear speaker and the ear.
To account for this variability, we collected paired recordings from human wearers across various environments. Participants wore our glasses prototype along with temporary in-ear microphones, providing simultaneous frame-microphone and ground-truth in-ear signals for model fine-tuning. 
Participants were free to move their head and body during recording to capture natural variability in pose.
To further adapt to each individual's anatomy at inference time, the system performed a brief calibration procedure in which a known audio stimulus was played through the glasses speakers to estimate the user-specific secondary path. This per-user estimate was then used to condition the neural network during inference. 

We evaluate the real-world ANC performance of our system on 11 unseen participants across 8 environments, with and without user-specific secondary path calibration. Our experimental design incorporated head shape variability, variations in glasses fit, and diverse environmental conditions. All environments were neither used for the collection of the mannequin head dataset, nor the human training dataset. We randomly placed the noise source in the environment and played unheard noise.

\noindent \textit{Noise reduction performance across users.} 
Fig.~\ref{fig:usereval}a shows our noise reduction performance across different frequencies with and without the user-specific secondary path calibration, yielding an overall reduction of 11.2 and 9.6~dB respectively in the 100-1000~Hz operating frequency range. 
Fig.~\ref{fig:usereval}b shows the performance for each participant with and without the user-specific secondary path calibration. Incorporating user-specific secondary-path calibration improved the average noise reduction across all users by 1.6~dB, and reduced inter-user variability, as reflected in the standard deviation decreasing from 1.5~dB to 0.4~dB. Fig.~\ref{fig:usereval}c shows example secondary path impulse response estimations from different participants. The impulse responses differ noticeably in shape, amplitude, and delay across participants, reflecting how variations in ear shape and glasses positioning alter the speaker-to-ear acoustic coupling.

\begin{figure*}[!ht]
    \centering
    \includegraphics[width=1\linewidth]{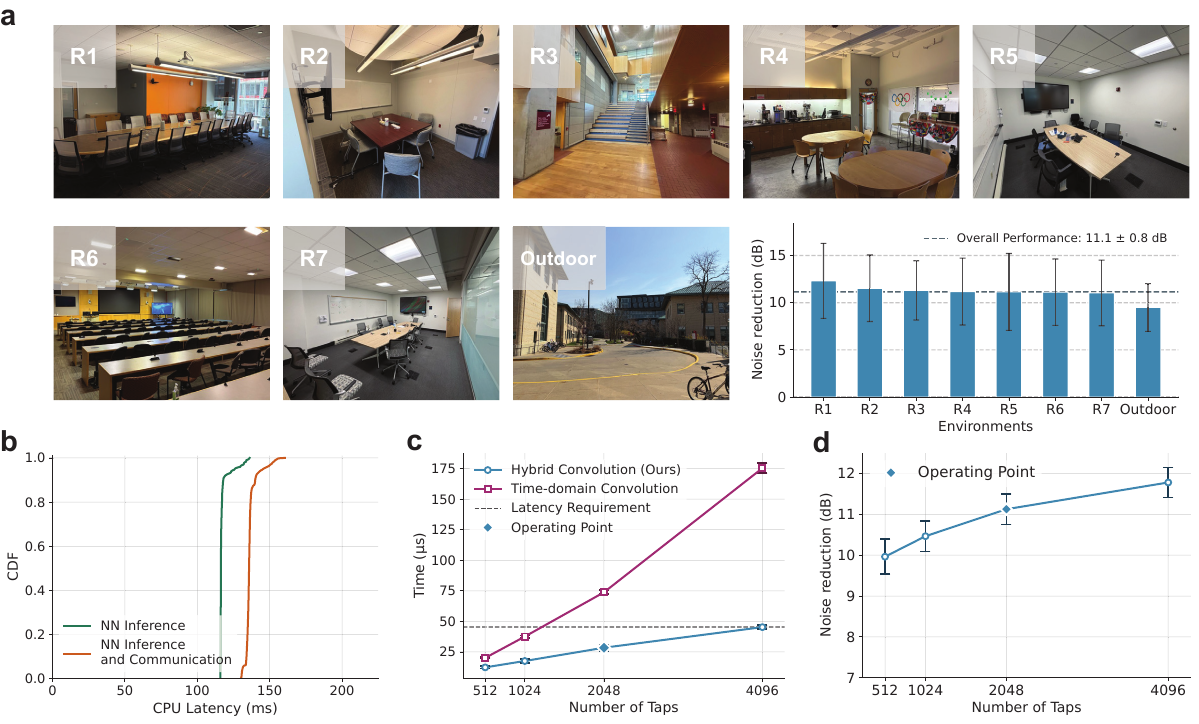}
    \caption{\textbf{Real-world evaluation across environments and system characterization. a,} \blue{Environments where our system was evaluated. These include conference rooms, a classroom, a kitchen, an atrium and an outdoor courtyard. See Extended Data Table \ref{tab:room_stats} for room characteristics. The bar plot shows the mean noise reduction for each environment.} \textbf{b,} CDF of neural network inference and communication latency on CPU to update transfer functions. We set our system to update the transfer functions every 200~ms. \textbf{c,} Processing time per audio block for time-domain convolution versus hybrid partitioned convolution, as a function of the number of taps per control filter. \blue{\textbf{d,} Noise reduction performance as a function of the number of taps per control filter estimated by the neural network.}}
    \label{fig:roomsruntime}
\end{figure*}

\noindent \textit{Noise reduction performance across environments.} 
Our evaluation spanned 8 different environments, including 7 indoor rooms and 1 outdoor space. \blue{The indoor rooms varied in size, layout, and material, all of which affect acoustic propagation and lead to a wide range of acoustic conditions. Specifically, their volumes range from approximately \qty{45}{\cubic\meter} in a small conference room to over \qty{960}{\cubic\meter} in a large lobby, and reverberation times measured by RT60 (the time it takes for the sound to decay by 60~dB) span about 0.48--0.90~s (see Extended Data Table~\ref{tab:room_stats} for details).} Fig.~\ref{fig:roomsruntime}a shows images of each of these environments. All indoor environments achieved a noise reduction of at least 11.0~dB, as shown in the bar plot of Fig.~\ref{fig:roomsruntime}a. \blue{Our performance drops in the outdoor environment due to the presence of wind, as wind is uncorrelated noise that is physically impossible to cancel at the eardrum, but our system still achieved a mean noise reduction of 9.5~dB.}

\noindent \blue{\textit{Noise reduction under glasses displacement.}}
\blue{In everyday use, the glasses position on the wearer's face may shift, causing the secondary path to deviate from the calibrated one. We evaluated the effect of these shifts with 6 participants performing four movements chosen to displace the glasses: looking down, shaking their heads, removing and re-wearing the glasses, and pedaling on an exercise bike. Each participant’s secondary path was calibrated once at the start of the first session and was not re-estimated at any point afterwards. Extended Data Fig. \ref{fig:glasses_displacement} shows that our noise reduction performance remained consistent despite any positioning changes of the glasses, which were 11.3, 11.2, 11.1 and 10.7 dB for the four conditions respectively, against 11.3 dB at rest. We attribute this to the long wavelengths of our target cancellation frequencies, where small shifts in the secondary path lead to only small phase errors.}

\noindent \textit{Audio playback clarity enhancement.} We next evaluated how our ANC system improves the intelligibility and signal fidelity of desired audio played back through the glasses speakers in the presence of background noise. We played random speech and music samples obtained from online datasets through the glasses speaker under identical noise conditions, with and without ANC enabled. Performance was quantified using short-time objective intelligibility (STOI), word error rate (WER), and scale-invariant signal-to-noise ratio (SI-SNR) (Fig.~\ref{fig:usereval}d).

Enabling ANC improved the intelligibility of speech played through the glasses speakers, increasing STOI from 0.59 to 0.68. Word error rate decreased across both automatic speech recognition models~\cite{whisperasr} evaluated: from 18.1\% to 11.0\% for Whisper Small, and from 12.0\% to 8.4\% for Whisper Medium. The improvement was larger for the smaller model, whereas the larger model was more robust to background interference. We further observed a mean SI-SNR improvement of 6.4~dB for speech (from $-$6.2 to 0.2~dB) and 5.7~dB for music (from $-$4.7 to 1.0~dB). \blue{As a reference, replaying the same speech without background noise yielded a STOI of 0.88, WER of 4.0\% and 3.6\% for Whisper Small and Whisper Medium, respectively. These results confirm that our ANC system not only reduces ambient noise but also enhances the clarity and fidelity of desired audio playback through the open-ear speakers, recovering a substantial portion of the intelligibility lost to ambient noise. A gap to the noise-free reference remains, which we attribute in part to our system’s focus on the low-frequency range, where residual noise above the operating band continues to mask the higher-frequency content of speech and music.}

\subsection*{Subjective user experience}
In addition to the objective evaluation, we conducted a user study to assess the perceived clarity of target audio and the intrusiveness of environmental noise. Each participant completed 10 trials with randomly selected speech (7 trials) or music (3 trials) clips.  Environmental noise was played through an external noise source while the desired audio was delivered through the glasses' open-ear speakers. To reduce bias, participants first heard the noise-only condition without ANC, followed by the noise mixed with speech or music played through either our glasses with ANC or the closed-ear device (AirPods 4), presented in randomized order. Participants rated audio clarity and noise intrusiveness on a 1--5 mean opinion score (MOS) scale (detailed in \nameref{sec:methods}).

Enabling our ANC system improved perceived audio clarity (Fig.~\ref{fig:usereval}e), with MOS increasing from 2.1 to 3.7 for speech and from 2.5 to 4.1 for music. Similarly, noise intrusiveness ratings (Fig.~\ref{fig:usereval}f) improved from 1.5 to 3.3 for speech and from 1.8 to 3.6 for music. 
For reference, AirPods 4 achieved clarity scores of 4.4 and 4.5, and intrusiveness scores of 3.7 and 4.1, for speech and music, respectively.
We note that AirPods 4 is included as a qualitative reference only to illustrate ANC performance of a closed-ear design.

\subsection*{Optimizations for real-time operation}
We evaluated the latency of different components of our system. Fig.~\ref{fig:roomsruntime}b shows the cumulative distribution function (CDF) of ANC filter update latency, which includes neural network inference and communication overhead. The median latency is \qty{136}{\milli\second} with a 95th percentile of \qty{145}{\milli\second}. Based on these values, we set the filter update period to \qty{200}{\milli\second}, providing sufficient margin for reliable real-time operation.

We next compared the per-sample processing time of direct time-domain convolution against our hybrid partitioned convolution as the number of ANC filter taps increases. At our sample rate of $f_s=$~\qty{22050}{\hertz}, each sample must be processed within \qty{45}{\micro\second} ($1/f_s$) to maintain real-time operation. As shown in Fig.~\ref{fig:roomsruntime}c, direct time-domain convolution meets this budget for filters up to 1024 taps, whereas our hybrid convolution extends this to 2048 taps.

We evaluated how noise reduction performance scales with the number of ANC filter taps per channel. Fig.~\ref{fig:roomsruntime}d shows that performance improves from 10.0~dB at 512 taps to 10.5~dB at 1024 taps and 11.1~dB at 2048 taps, with a further increase to 11.8~dB at 4096 taps. We select 2048 taps as our operating point, as it achieves strong noise reduction while operating well within the real-time latency constraint of the hybrid convolution scheme.

We also investigated the effect of DSP computation latency on noise reduction performance (Extended Data Fig.~\ref{fig:delay_noise_reduction}). We simulated different latency values including $1/f_s$, $2/f_s$, $4/f_s$, $8/f_s$, and $16/f_s$ (corresponding to 45--726~\unit{\micro\second}). Noise reduction degrades as latency increases: from 11.1~dB at \qty{45}{\micro\second} to 4.4~dB at \qty{726}{\micro\second}. This confirms that minimizing DSP latency is critical for effective ANC, as additional processing delay introduces phase errors between the anti-noise and the incoming noise waveform that degrade destructive interference.

\section*{Conclusions}

We present an active noise cancellation system for open-ear smart wearables that operates without relying on a sealed ear canal or an inward-facing error microphone. The proposed virtual in-ear sensing framework enables the system to estimate the acoustic signal at the ear from microphones distributed across the frame, while the hybrid neural-DSP pipeline satisfies the latency requirements of real-time anti-noise generation. 

\blue{Our system attenuates the 100–1000 Hz band in which environmental noise is concentrated. It also lacks any passive isolation, which would otherwise attenuate high-frequency sounds. Higher-frequency content therefore passes largely unattenuated, including much of the information-bearing energy of speech and the tonal components of alarms and sirens, so the system reduces intrusive background noise while leaving these cues audible.} 

\blue{Beyond improving audio playback in noisy environments, this capability lays the groundwork for selective auditory experiences rather than all-or-nothing noise blocking, letting users quiet unwanted noise while staying aware of the sounds that matter to them. Prior work on spatially selective ANC~\cite{xiao2023spatially, xiao2024icassp, Xiao2025fa, Xiao2025soft}, semantically aware listening~\cite{veluri2023semantic}, and personalized sound zones~\cite{Chen2024soundbubble} points toward such experiences, but all of it rests on an ANC capability that open-ear devices have lacked. Our system provides the missing foundation on which such selective systems are built for open-ear devices, making these interfaces achievable in exactly the settings, such as conversation or street navigation, where occluding the ear is impractical or unsafe.}


{We note that recent commercial earbuds have begun offering noise reduction in non-occluding form factors that do not seal the ear canal~\cite{airpods4, shokzopenfit}.
However, these devices still rely on an inward-facing microphone positioned at the entrance of the ear canal opening to capture a local acoustic reference for ANC feedback.
In contrast, our system uses only microphones on the glasses frame and a neural network to virtually estimate the in-ear signal for feedback-free open-ear ANC.}

We demonstrate that the underlying approach of our system applies in principle to broad classes of open-ear wearables beyond smart glasses such as smart headsets, headbands and helmets. Indeed, augmented and virtual reality headsets such as Meta Quest~\cite{metaquest3}, Microsoft HoloLens~\cite{hololens2}, and Apple Vision Pro~\cite{applevisionpro} are increasingly deployed in professional training contexts such as surgical simulation, military field exercises, and industrial skills training. Importantly, these devices have larger form factors that distribute microphones across the head-mounted display at varying distances from the ear, which can increase the acoustic lookahead available for certain source directions and potentially enable more effective cancellation on these platforms.

{While our prototype uses eight microphones (four per side), current commercial smart glasses typically incorporate fewer---for example, the Ray-Ban Meta glasses feature five microphones~\cite{raybanmeta} and the Meta Aria Gen 2 glasses feature seven~\cite{aria}.
Our microphone ablation study (Fig.~\ref{fig:mannequin}g, Extended Data Fig.~\ref{fig:mic_config}) shows that the system retains effective noise reduction performance with fewer microphones, which suggests that the proposed framework could be deployed on existing or earlier-generation open-ear wearable devices without requiring significant hardware modifications.}

However, we highlight key limitations of our current prototype. First, while our model generalizes across users and rooms, the user-specific secondary path calibration still improves performance relative to no calibration. The calibration is currently performed once with a temporary in-ear microphone when the user puts on the glasses for the first time. Our evaluation across 11 users with diverse head shapes and glasses fits suggests the system is robust to substantial anatomical variability. \blue{Although we have demonstrated that our system is robust to regular user movements, secondary path calibration accuracy may still degrade in conditions our evaluation does not cover. Objects introduced between the speaker and the ear, such as hair, a hat or a helmet, may alter the acoustic propagation, and transducer sensitivity may drift over long term wear \cite{mems_drift}. In both cases, an outdated secondary path calibration may lead to degraded cancellation performance. We believe that future research into periodic lightweight recalibration using the transfer function between speakers and frame microphones could mitigate this going forward.}

\blue{Our system's performance decreases in outdoor environments, from a mean of 11.3 dB indoors to 9.5 dB in the outdoor courtyard, primarily due to wind. Wind is a known challenge for ANC in general: turbulent airflow acting on the microphone membranes is not a propagating acoustic field, so it produces uncorrelated pressure fluctuations that the system misinterprets as ambient sound and for which no correct anti-noise exists \cite{chen2024effect}. Commercial devices mitigate this by reducing the feedforward gain when wind is detected, identifying it from segments of high energy and low coherence between reference microphone pairs \cite{bosewind, applewind, applewind2}. The same cues are available on our platform, and gating filter updates during detected wind would prevent turbulence from corrupting the estimated filters.}

\blue{Our subjective user study was not fully blinded. Participants could tell when they were listening through the AirPods because open-ear glasses and earbuds are physically distinguishable to the wearer. We include the AirPods result as a qualitative reference rather than a controlled equivalent for comparison. This limitation only exists for the AirPods comparison, as the glasses ANC-on and ANC-off ratings used identical hardware under unlabeled conditions, leaving them unaffected by any listener bias.}

\blue{Our DSP generates anti-noise continuously with an end-to-end latency of \qty{113}{\micro\second}, while the neural network updates the ANC filters every \qty{200}{\milli\second}. As long as the acoustic scene is approximately stationary within the \qty{200}{\milli\second}, the filters remain valid. Rapid changes in the acoustic environment, such as fast head movements or quickly moving acoustic sources, are therefore limited by our neural network filter update latency, and filters computed for an earlier geometry provide reduced cancellation until the next update. We note that this update period is set by our current inference and communication latency rather than by the approach itself, and can be reduced through further architectural optimizations. Our current network processes the full \qty{2}{\second} input window with overlapping segments and recomputes the entire output at each update. Redesigning the network to incrementally process only the latest audio segment would reduce the per-update computation. Additionally, migrating inference from a general-purpose CPU to a dedicated neural processing unit, such as the Hexagon NPU available on the Snapdragon AR1 platform, would further accelerate execution. Finally, the inter-processor communication in our prototype relies on User Datagram Protocol (UDP) over an Ethernet cable between two separate boards. In an integrated System-on-Chip (SoC), this overhead would be eliminated by shared-memory transfers between the application processor and the DSP subsystem. Together, these optimizations could reduce the filter update period to enable faster adaptation to dynamic acoustic scenes.}



Our prototype uses two off-the-shelf single-board processors: a Raspberry Pi 5 (Broadcom BCM2712, quad-core Cortex-A76 at 2.4 GHz) for neural network inference and a Bela board (BeagleBone Black with a 200 MHz TI AM335x programmable real-time unit) for ultra-low-latency DSP processing.
Together, these boards consume approximately 5–7 W in our experiment, giving a battery life of around 4 hours when connected to a 5~V, 5000~mAh battery.
Importantly, this dual-processor architecture---separating neural inference from deterministic real-time filtering---mirrors the compute organization already present in commercial smart glasses SoCs such as the Qualcomm Snapdragon AR1 Gen 1 used in the Ray-Ban Meta glasses.
Transitioning our system to such an integrated SoC would substantially reduce power consumption, consolidate the hardware into a smaller form factor, and enable further latency optimizations for inference and communication.


\section*{Methods} \label{sec:methods}

\subsection*{Study design}
This study was approved by Carnegie Mellon University’s Institutional Review Board \\ (STUDY2025\_00000146). All studies complied with relevant ethical regulations. Participants were recruited by word of mouth from the Carnegie Mellon University student community. Written consent was obtained for human subjects participating in the study. Randomization was not applicable and investigators were not blinded. Participants above the age of 18 were eligible for the study.

\subsection*{Hardware Setup}

\header{Glasses prototype.} We designed a custom 3D-printed glasses prototype. We embedded 8 MEMS microphones (SPH8878LR5H-1) in the frame of the glasses to capture spatial acoustic information (see Fig.~\ref{fig:concept}b). We designed a custom breakout board for 6 of the microphones that retained the same circuit schematic as the off-the-shelf Sparkfun board but with a modified PCB layout to fit within the narrow edges of the 3D-printed glasses frame (Extended Data Fig.~\ref{fig:mic_pcb}).
The remaining 2 microphones used the off-the-shelf Sparkfun breakout board. \blue{For our anti-noise speakers on the glasses frame we used the speaker components taken from a disassembled pair of Meta Oakley Vanguard glasses and connected them to an Adafruit MAX98306 stereo audio amplifier. Although these are open-ear speakers, the acoustic crosstalk from the speaker to the opposite ear is minimal due to the directionality of the speaker, and the longer acoustic path that attenuates the contralateral signal. We measured 33.1 dB more broadband energy and 32.3 dB more energy in the 100-1000 Hz band at the ipsilateral ear than the contralateral ear.}

\header{Microcontrollers.} We used a Bela cape Rev B with BeagleBone Black board and the Bela Audio Expander Capelet as the low-latency DSP board \cite{bela1}. It handled the recording, processing, and playing of audio signals. \blue{We connected all 8 MEMS microphones to the analog inputs of the board, which are routed through the Audio Expander Capelet. The Audio Expander Capelet provides analog anti-aliasing filtering, DC blocking and level shifting on every channel. The speakers and amplifiers were connected to the analog outputs. This was to take advantage of the lowest latency path on the Bela board. We recorded the ground truth signals by connecting the in-ear microphones to the Bela audio inputs. We set the sampling rate of the recorded signals to the lowest supported rate at 8 channels, \qty{22050}{\hertz}}. 

The DSP board used a deterministic block-based processing paradigm, where each audio block was captured, processed, and output within a fixed time window set by the sampling rate and block size. For the real-time audio thread, the computation must be completed before the next block arrives, otherwise an underrun occurs. Consequently, the computation itself introduces zero additional processing latency. We set the audio block size to be the smallest possible value to minimize latency, which is 1 sample at \qty{22050}{\hertz}, meaning that each audio block has \qty{45}{\micro\second} for processing. \blue{We measured the end-to-end loopback latency from microphone input to speaker output to be \qty{113}{\micro\second}, consisting of \qty{45}{\micro\second} of computation and \qty{68}{\micro\second} of the remainder of the chain, comprising the analog anti-aliasing and signal-conditioning stages, the ADC and the DAC.}

\blue{The choice of sample rate is driven by the trade-off between achievable latency and real-time computation budget. A lower sample rate would be attractive computationally, since a fixed impulse-response duration would be covered by fewer taps, and our 100–1000 Hz operating band does not itself require a 22,050 Hz sampling rate. However, the sample period sets a lower bound on achievable latency where even with single-sample blocks the per-sample budget would be at least 1$/f_s$. For example, a sample rate of 8 kHz would raise the computation latency from 45 µs to 125 µs. Since cancellation degrades steeply with added delay (Extended Data Fig. 6), the saving in computation would cost more in latency than it recovers. Conversely, a higher sampling rate reduces the per-sample computational budget. At 44,100 Hz the budget halves to 22.7 µs, which our four channel hybrid convolution cannot meet, and spanning the same impulse response duration would require twice the taps. We therefore operate at 22,050 Hz, the lowest sampling rate supported by our hardware, which keeps the latency floor at 45 µs while leaving sufficient budget for real-time filtering.}

We deployed the neural network on a Raspberry Pi 5. We recorded the microphone signals simultaneously on the Raspberry Pi and the DSP board to reduce the communication overhead of audio samples from the DSP to the Raspberry Pi for neural network input. We attached an ADC8x HAT to the Raspberry Pi for recording. We connected the clock pins from the DSP board onto the Raspberry Pi, such that both boards follow the sampling clock. A calibration was performed once by playing white noise while both boards sampled from the same microphone in order to compensate for the magnitude response differences between the two ADCs.

We connected the DSP board to the Raspberry Pi using the Ethernet ports. The Raspberry Pi performed neural network inference, and also calculated the FFT blocks for the partitioned segments of the ANC filters, transmitting them to the DSP board. Communication was done via UDP to minimize communication overhead. The latency of this communication and the neural network inference are shown in Fig. \ref{fig:roomsruntime}b.

\header{Ground truth measurements.} \blue{Our ground truth is the binaural acoustic signals measured at the user's ears. To obtain the measurements, we used a set of Soundman OKM II Studio in-ear binaural microphones that the user wore throughout the data collection and evaluation stages. These are blocked-canal style microphones that rest at the entrance of the ear canal. Sound transmission from the entrance to the eardrum is a fixed, direction-independent filter \cite{transmission_earcanal, fundamentals_acoustic}, so the entrance retains the direction-dependent component of the field that our system estimates, and reducing the residual pressure there reduces it correspondingly at the eardrum. Additionally, the signals in our 100 to 1000 Hz operating band are well below the ear canal's first resonance, and have relatively long wavelengths compared to the canal length, making the measurement insensitive to small variations in placement depth. These microphones were connected to the audio inputs of the Bela board. Since the Bela samples the audio channels at a fixed sample rate of 44,100 Hz, we downsampled the recorded signals offline for training and evaluation.}

\header{Primary source loudspeaker.} \blue{We utilized a JBL Flip 5 bluetooth speaker to play the various ambient noises when collecting our mannequin head and user data. This allowed us to easily collect noise recordings from multiple environments, as well as vary the noise source distance, elevation and azimuth relative to the glasses frame. To demonstrate that the specific hardware of the noise source has little impact on the noise reduction performance of our system, we chose 3 speakers at varying price and quality. Specifically, we additionally used the Genelec 8010, and the Music Legend T18 speakers. The results of our additional experiment can be found in the Results section, and Extended Data Fig. \ref{fig:source_speaker}.} 

\subsection*{ANC primer}

We first described the basic formulation of a conventional multiple-input single-output (MISO) feedforward ANC system. As described above, the eight microphones and two speakers on the glasses are divided into two independent groups, one per ear, with four microphones and one speaker per side. The following formulation describes one such side. We considered a set of $M=4$ frame microphones as the reference microphones distributed along the frame of the glasses that measured the incoming noise, a speaker that emitted an anti-noise signal, and an error microphone placed near the entrance of the ear canal that measured the residual noise. The objective was to minimize the residual noise power at the error microphone, which approximated the sound perceived by the user.

Let $X_m(z)$ be the $z$-domain representation of the noise signal captured by the $m$-th frame microphone. As the noise \blue{propagates} to the error microphone, its spectral and temporal characteristics are filtered by the primary acoustic path $P_m(z)$ between the $m$-th frame microphone and the error microphone. The primary noise reaching the ear in the $z$-domain is given by $D(z) = \sum_{m=1}^{M} P_m(z)X_m(z)$. To achieve noise reduction, the system utilizes a set of $M$ ANC filters $W_m(z)$ to generate the speaker drive signal $Y(z) = \sum_{m=1}^{M} W_m(z)X_m(z)$. The drive signal $Y(z)$ is propagated through the secondary path $S(z)$, which represents the complete electro-acoustic chain, including the DAC, ADC, amplifiers, speaker response, and physical acoustic propagation from the speaker to the ear. We denote the anti-noise signal reaching the ear as $\hat{D}(z) = S(z)Y(z)$. The residual error signal $E(z)$ at the error microphone is the superposition of the primary noise and the anti-noise
\begin{subequations}
    \begin{align}
        E(z) &= D(z) + \hat{D}(z) \label{eq:E_z_sum} \\
             &= D(z) + S(z)Y(z) \label{eq:E_z_D} \\
             &= \sum_{m=1}^{M} P_m(z)X_m(z) + S(z) \sum_{m=1}^{M} W_m(z)X_m(z). \label{eq:E_z_P}
    \end{align}
\end{subequations}
The optimal filter for each channel $m$ that satisfied $E(z) = 0$ is theoretically given by
\begin{equation}
    W_{m,\, \mathrm{opt}}(z) = -\frac{P_m(z)}{S(z)},
\end{equation}
where the subscript $(\cdot)_\mathrm{opt}$ denotes an optimal solution.

In practice, the system operated in the time domain with FIR filters of length $L_C$. We let $d[n]$ denote the primary noise reaching the ear, and $\hat{s}[n]$ is the estimated impulse response of the secondary path. The speaker drive signal $y[n]$ is generated by convolving the reference signals $x_m[n]$ with their respective filters $w_m[n]$ as
\begin{equation}
    y[n] = \sum_{m=1}^{M} (w_m[n] * x_m[n]),
    \label{eq:y_n}
\end{equation}
where $*$ denotes time-domain convolution. We denote the anti-noise signal reaching the ear as $\hat{d}[n]$, which is expressed as the drive signal convolved with the estimated secondary path as
\begin{equation}
    \hat{d}[n] = \hat{s}[n] * y[n] = \hat{s}[n] * \sum_{m=1}^{M}(w_m[n] * x_m[n]) .
    \label{eq:d_hat_n}
\end{equation}
The time-domain residual error $e[n]$ is 
\begin{equation}
    e[n] = d[n] + \hat{d}[n].
    \label{eq:e_n_sum}
\end{equation}
To account for the secondary path during filter optimization, a filtered-reference signal is typically defined as $r_m[n] = \hat{s}[n] * x_m[n]$. Using the commutativity of convolution, equation~\eqref{eq:e_n_sum} can be rewritten in vector form as
\begin{equation}
    e[n] = d[n] + \sum_{m=1}^{M} \mathbf{w}_m^{\mathrm{T}} \, \mathbf{r}_m[n],
    \label{eq:e_n_vec}
\end{equation}
where $\mathbf{w}_m = [w_m[0], w_m[1], \dots, w_m[L_C-1]]^{\mathrm{T}}$ is the filter coefficient vector and $\mathbf{r}_m[n] = [r_m[n], \allowbreak r_m[n-1], \dots, r_m[n-L_C+1]]^{\mathrm{T}}$ is the filtered-reference signal vector. By concatenating these vectors for all $M$ channels into $\mathbf{w} = [\mathbf{w}_1^{\mathrm{T}}, \dots, \mathbf{w}_M^{\mathrm{T}}]^{\mathrm{T}}$ and $\mathbf{r}[n] = [\mathbf{r}_1^{\mathrm{T}}[n], \dots, \mathbf{r}_M^{\mathrm{T}}[n]]^{\mathrm{T}}$, we defined a cost function $J(\mathbf{w})$ to minimize the expected squared error while constraining the filter energy to prevent overloading the speaker as
\begin{equation}
    J(\mathbf{w}) = \mathcal{E}\{e^2[n]\} + \beta \mathbf{w}^\mathrm{T} \mathbf{w},
\end{equation}
where $\mathcal{E}\{\cdot\}$ denotes the expectation operator, and $\beta$ denotes the regularization factor. The optimal time-domain solution is then given by:
\begin{equation}
    \mathbf{w}_\mathrm{opt} = -(\bm{\Upphi}_{\mathbf{rr}} + \beta \mathbf{I})^{-1} \bm{\upphi}_{\mathbf{r}d},
\end{equation}
where $\bm{\Upphi}_{\mathbf{rr}} = \mathcal{E}\{\mathbf{r}[n]\mathbf{r}^{\mathrm{T}}[n]\}$ is the $(M \cdot L_C) \times (M \cdot L_C)$ autocorrelation matrix of the filtered-reference signals, $\mathbf{I}$ denotes the identity matrix, and $\bm{\upphi}_{\mathbf{r}d} = \mathcal{E}\{\mathbf{r}[n]d[n]\}$ denotes the $(M \cdot L_C) \times 1$ cross-correlation vector between the filtered-reference signals and the primary noise. 

\header{Wiener filter baseline.} The closed-form solution $\mathbf{w}_\mathrm{opt}$ derived above served as the Wiener filter solution used in our evaluation (Extended Data Table \ref{tab:wiener}). Computing this solution required the physical in-ear error microphone signal $d[n]$ to form the cross-correlation vector $\bm{\upphi}_{\mathbf{r}d}$, which was precisely the signal unavailable in an open-ear system during deployment. For each evaluation segment, the Wiener filter was estimated from the same $L_N$-length context window used as the neural network input and produced a filter of the same $L_C$-tap length, ensuring a fair comparison. We computed the autocorrelation matrix $\bm{\Upphi}_{\mathbf{rr}}$ and the cross-correlation vector $\bm{\upphi}_{\mathbf{r}d}$ from the recorded reference and in-ear microphone signals over this window, and solved for $\mathbf{w}_\mathrm{opt}$ in closed form. \blue{We set the regularization term to be}

\begin{equation}
    \blue{\beta = \beta_0 \cdot \frac{\mathrm{tr}(\bm{\Upphi_{\mathbf{rr}}})}{M \cdot L_C}}
\end{equation}
\blue{where $\beta_0 = 10^{-4}$. This regularization is -40 dB relative to the average power of the filtered reference signals, and is -57 dB to -65 dB relative to the largest eigenvalue across evaluation segments.}

\blue{This baseline represented the best achievable performance of an approach that had access to the ground-truth ear signal, acting as an oracle upper bound for our neural network. Throughout our experiments, the Wiener filter was applied using the hybrid partitioned convolution method, the exact same way our network predicted control filters were applied. This keeps the comparisons between the two consistent. The two filters used the same 2048 tap length, secondary path model, application method and end to end latency. The only difference was the way these 2048 taps were obtained.}

\header{Secondary path estimation and calibration.} Different users have different head anatomies and therefore different secondary paths. To account for this variation, we modeled the secondary acoustic path $\hat{s}[n]$ as a 1024 tap FIR filter. We estimated this path by playing an exponential sine sweep (ESS) from \qty{20}{\hertz} to \qty{11025}{\hertz} over 20 seconds from the glasses speaker and recorded the output from the in-ear microphone. This was done twice, and the two recorded signals were averaged. We then calculated the corresponding inverse filter and used it to deconvolve the averaged recorded signal to recover the secondary path impulse response in the time domain \cite{farina2000sweptsine}. To compensate for the frequency response of the speaker we applied a first order magnitude shaping to the excitation sweep and the inverse filter.

\blue{The ESS is a standard method for impulse response measurement as it is highly robust even in noisy environments. Consequently, calibration does not require a quiet environment. As shown in Extended Data Fig. \ref{fig:calibration_snr}, the performance of our system stays consistent when the secondary path is calibrated across background-noise levels from 30 down to 0 dB SNR, and degrades by less than 1.7~dB at -20 dB SNR. Here, calibration SNR denotes the level of the sweep stimulus relative to the ambient noise measured at the in-ear microphone.}

\blue{Since the secondary path model is only used during training and as an input to the neural network, it is not on the latency critical real-time audio path. Therefore, we chose 1024 taps conservatively to ensure the model fully captures the acoustic path. We also verified that shorter secondary path models are sufficient. Truncating the estimated secondary path impulse response and using it without retraining the network yields noise reduction within 0.1~dB of the full 1024-tap model down to 32 taps, which is roughly the duration of the physical impulse response.}

The secondary path estimation and calibration was performed once for each user before data collection and real-time ANC operation. This secondary path estimation was used when calculating the loss during our network training step (Eq.~\eqref{eq:d_hat_n}). The path is also included as an input to our neural network (details below). When the calibrated secondary path was not available during inference, we used an averaged secondary path over all previous secondary path estimations we obtained as input instead.

\header{Acoustic feedback cancellation (AFC).} The anti-noise signal emitted by the speaker was partially captured by the frame microphones, causing the acoustic feedback that may cause system instability (commonly perceived as whistling or howling). We employed a standard feedback cancellation technique~\cite{kuo1999active} to eliminate this effect. Similar to the secondary path modeling, the feedback path from the speaker to the $m$-th frame microphone was characterized as an estimated FIR filter $\hat{g}_m[n]$. A filter length of 256 taps was chosen to sufficiently capture the short physical path while minimizing computational overhead. These paths were estimated for each of the $M=4$ frame microphones using the same ESS method used for secondary path estimation.

Once these paths were estimated, they were integrated into the real-time ANC loop to perform feedback cancellation. At each timestamp $n$, for each microphone $m$, the system convolved the speaker drive signal from the previous timestamp $y[n-1]$ with the corresponding estimated leakage path $\hat{g}_m[n]$ to estimate the feedback signal received at that microphone. This estimate was then subtracted from the raw microphone signal $x'_m[n]$ to recover the clean reference noise signal $x_m[n]$ used for both the neural network inference and the real-time ANC filtering. Crucially, this $x_m[n]$ was the clean signal used to generate the speaker drive signal $y[n]$ and the resulting estimated anti-noise $\hat{d}[n]$ established in equation~\eqref{eq:y_n}, i.e.,
\begin{equation}
    x_m[n] = x'_m[n] - (y[n-1] * \hat{g}_m[n]).
    \label{eq:feedback_cancel}
\end{equation}
By implementing the AFC algorithm, it ensured that the ANC filters were calculated based on the external environmental noise signals rather than the system's own output, thereby maintaining system stability.

\subsection*{Neural network architecture}

\header{Problem definition.} Unlike conventional systems that relied on a physical error microphone to compute $\mathbf{w}_\mathrm{opt}$, the open-ear system performed virtual in-ear sensing. The goal of our neural network $f$ was to map the noise signals $x_m[n]$ captured by the $M=4$ frame microphones to the corresponding set of causal control filters $w_m[n]$ that minimized the error at the user's ear:
\begin{equation}
    f(x_1[n], \dots, x_M[n]) = \{w_1[n], \dots, w_M[n]\}.
\end{equation}
These estimated filters were used to generate the anti-noise signal in real-time. As formulated in equation~\eqref{eq:d_hat_n}, the anti-noise interacted with the physical secondary path $s[n]$ before reaching the user's ear. The neural network was designed to continuously estimate the optimal filters such that the magnitude of the residual error $e[n]$ was minimized across dynamic acoustic environments without the need for a persistent physical error signal.

\header{Loss function.} We trained the model using an $L_1$ loss function based on the residual error at the ear. Using the estimated anti-noise signal $\hat{d}[n]$ as defined in equation~\eqref{eq:d_hat_n}, the loss function was defined as the mean absolute error of the residual signal:
\begin{equation}
    \mathcal{L} = \| e[n] \|_1= \| d[n] + \hat{d}[n] \|_1 ,
\end{equation}
where $d[n]$ was the ground-truth primary noise measured at the user's ear during training. By minimizing this loss, the neural network learned to estimate filter coefficients $\mathbf{w}$ that minimize the residual noise at the ear without requiring a physical error microphone during real-time inference.

\blue{Although the system objective is to minimize acoustic energy at the ear, which would suggest an $L_2$ training objective, we found $L_1$ to perform better on that energy metric. Otherwise identical models trained with $L_1$ and $L_2$ losses achieved 10.4~dB and 9.7~dB of noise reduction respectively, in the 100-1000~Hz band. We attribute this to how the two losses allocate model capacity across the residual spectrum. The $L_2$ gradient weights each error sample by its own magnitude, so training is dominated by the largest residual components. In our case these are components no causal filter can cancel: energy below the secondary path's low-frequency roll-off, and high-frequency energy where the frame microphones and the ear are weakly coherent. $L_1$ weights errors more uniformly, directing more capacity toward the band in which cancellation is achievable.}

\header{Encoder and feature extractor.} The input microphone signals, originally sampled at \qty{22050}{\hertz}, were first decimated to \qty{8820}{\hertz} to reduce the computational cost of subsequent processing, since active cancellation operated primarily in the low-frequency range.
An STFT with an FFT size of 1024 and a hop length of 256 was applied to each channel, producing $F = 513$ frequency bins per frame.
From the STFT representations, we extracted three complementary feature types with respect to a designated reference microphone $X_\mathrm{ref}(t,f)$ (chosen as the microphone closest to the speaker).
The reference spectrogram contributed its real and imaginary parts as 2 input channels.
The interchannel phase difference (IPD) was computed for each of the $M-1 = 3$ non-reference microphones as:
\begin{equation}
    \text{IPD}_m(t,f) = \bigl[\sin(\angle X_\mathrm{ref}(t,f) - \angle X_m(t,f)),\; \cos(\angle X_\mathrm{ref}(t,f) - \angle X_m(t,f))\bigr],
\end{equation}
yielding $2 \times 3 = 6$ channels that encoded the spatial phase relationships between microphones.
The interchannel level difference (ILD) was computed as:
\begin{equation}
    \text{ILD}_m(t,f) = \log_{10}\!\left(\frac{|X_\mathrm{ref}(t,f)| + \epsilon}{|X_m(t,f)| + \epsilon}\right),
\end{equation}
where $\epsilon$ was a small constant for numerical stability, yielding 3 additional channels.
In total, the encoder produced $2 + 6 + 3 = 11$ input feature channels per time-frequency frame, stacked along the channel dimension and passed to the subsequent network.

\header{Network architecture.} The feature tensor was processed by a U-Net encoder--decoder with skip connections.
The encoder consisted of 7 convolutional layers with channel progression $[32, 64, 64, 128, 128, 256, 256]$, each comprising a Conv2d with kernel size $(5, 2)$ and stride $(2, 1)$, followed by batch normalization and PReLU activation. Each encoder stage halved the frequency dimension while preserving the time dimension.
On each skip connection, a squeeze-and-excitation (SE) block~\cite{hu2018squeeze} performed channel-wise recalibration with a reduction ratio of 4: global average pooling compressed the spatial dimensions, two fully connected layers modeled channel interdependencies, and a sigmoid gate rescaled the feature channels before concatenation with the decoder.
At the bottleneck, a single-layer unidirectional LSTM with a hidden size of 128 processed the flattened features along the time axis to capture temporal dependencies across successive estimation windows.
The decoder mirrored the encoder with transposed convolutions, each concatenating the SE-recalibrated skip features before upsampling.
A final convolutional layer projected the decoder output to $2M = 8$ channels, representing the real and imaginary parts of the frequency-domain ANC filter for each of the $M = 4$ microphone channels.
\blue{Within each forward pass, the decoder outputs one filter estimate per STFT time frame. These per-frame estimates were averaged into a single filter for estimation stability.} 
The model contained 4.2M parameters and required 5.3G MACs per inference.

An ablation study (Extended Data Table~\ref{tab:ablation}) showed that the reference channel input was the most critical component: removing it degraded noise reduction from 10.4 to 9.1~dB, and removing it together with the LSTM further reduced performance to 9.0~dB.
Removing the LSTM or SE blocks individually each yielded 10.3~dB.
We also varied model size: a Small variant (1.1M parameters, 1.4G MACs) achieved 9.7~dB and a Large variant (9.4M parameters, 11.8G MACs) achieved 10.6~dB; the Base configuration was selected as it offered a good trade-off between noise reduction and computational cost.

\header{Secondary path conditioning.} To condition the network on user-specific speaker-to-ear acoustics, the estimated secondary path impulse response $\hat{s}[n]$ was truncated or zero-padded to a fixed length of 1024 taps.
A real-valued FFT was applied, and the real and imaginary parts of the resulting 513 frequency bins were concatenated and flattened into a $1026$-dimensional vector.
A multilayer perceptron (MLP) with hidden dimensions $[256, 128]$ compressed this vector into a 128-dimensional embedding.
A feature-wise linear modulation (FiLM)~\cite{perez2018film} generator then produced affine parameters $(\gamma, \beta)$ from this embedding, which modulated the LSTM bottleneck features as $\gamma \odot \mathbf{h} + \beta$, where $\odot$ denoted element-wise multiplication.
This mechanism allowed the decoder to adapt its filter predictions to each user's ear acoustics without re-training.

\header{Filter shaping.} The decoder output represented frequency-domain filters at the reduced sample rate of \qty{8820}{\hertz} with an FFT size of 1024.
The predicted filters were first averaged across time frames for estimation stability.
A half-cosine taper was applied to the top 12.5\% of frequency bins to suppress aliasing artifacts near the Nyquist frequency.
The tapered spectrum was then zero-padded to restore the full filter length of $L_C = 2048$ taps at the original sample rate of \qty{22050}{\hertz}.
After applying the inverse FFT, a 10\% time-domain fade-out window was applied to the tail of the filter to ensure smooth temporal decay.
The resulting time-domain FIR filters were transmitted to the DSP for real-time convolution.

\header{Real-time streaming.} During deployment, the neural network operated in a streaming fashion with an update interval of $L_D = \qty{200}{\milli\second}$.
At each update, the Raspberry Pi provided $L_N = \qty{2}{\second}$ of the most recent microphone signals as context to the model, which processed the STFT frames in a single forward pass and produced one set of $M = 4$ time-domain FIR filter coefficients.
The LSTM hidden state was carried across successive windows, enabling the model to maintain temporal coherence without reprocessing the full context from scratch.
The resulting filters were transmitted to the DSP board via UDP, where they immediately replaced the previous filter coefficients used in the hybrid partitioned convolution.
To avoid audible discontinuities at filter update boundaries, the DSP crossfaded between the old and new filter coefficients over the duration of one block ($B = 128$ samples).

\header{ONNX optimization.} We converted our PyTorch model into Open Neural Network Exchange (ONNX) opset 17 and used ONNX Runtime for inference on the Raspberry Pi 5.
Several operations that lacked efficient ONNX support were replaced with pure-tensor equivalents: the STFT was reimplemented as two Conv1d operations (cosine and sine filter banks), complex-valued arithmetic was decomposed into separate real and imaginary tensor paths, the FFT in the secondary path encoder was fused into the first MLP weight matrix as a single matrix multiplication, and the IFFT in the filter shaper was replaced with a precomputed inverse DFT matrix that folded in the frequency taper and upsampling.
ONNX Runtime leveraged ARM NEON kernels on the Raspberry Pi 5 for matrix multiplications and convolutions, yielding lower inference latency compared to PyTorch.

\header{Training.} We employed a two-stage training strategy with a context window of $L_N = \qty{2}{\second}$ in both stages.
In the pre-training stage, the model was trained on the mannequin dataset collected across three rooms (two rooms held out for evaluation).
We used the AdamW optimizer with a learning rate of $4 \times 10^{-4}$ and no weight decay, a batch size of 32, and the $L_1$ loss on the residual error.
The learning rate followed a constant-then-cosine schedule: it was held constant for 50 epochs and then annealed via cosine decay to $10^{-6}$ over the subsequent 50 epochs, for a total of 100 epochs.
In the fine-tuning stage, we resumed from the pre-trained checkpoint and fine-tuned on the real user dataset.
The learning rate was reduced to $10^{-4}$, held constant for 30 epochs, and then annealed via cosine decay to $10^{-6}$ over 20 epochs, for a total of 50 epochs.
All other hyperparameters were preserved from pre-training.
We used the last checkpoint for evaluation.
This two-stage strategy leveraged the larger mannequin dataset for robust initialization and then adapted to variability in human head geometry and glasses fit.

We organized human listeners into three groups for cross-group evaluation: Group~1 (5 users, 4 environments), Group~2 (5 users, 3 environments), and Group~3 (6 users, 5 environments).
The model was fine-tuned on Groups~1 and~2 and evaluated on Group~3, and separately fine-tuned on Groups~1 and~3 and evaluated on Group~2.
The reported noise reduction was averaged across both held-out groups.

To evaluate the model without user-specific secondary path calibration, we removed the secondary path encoder branch from the network and fine-tuned the resulting model for 20 epochs using a cosine learning rate schedule from $10^{-4}$ to $10^{-6}$.
As shown in Extended Data Table~\ref{tab:my_label}, the two-stage strategy (pre-training on mannequin data followed by fine-tuning on human data) achieved 11.2~dB noise reduction with calibration, compared with 10.4~dB when training on mannequin data alone or 10.1~dB when training on human data alone.
Without calibration, the two-stage model still achieved 9.6~dB, outperforming the mannequin-only model (7.6~dB) and the human-only model (9.1~dB).

\header{Data augmentation.} We applied several augmentation strategies during training to improve the model's robustness.
First, speed perturbation randomly resampled each training waveform by a factor drawn uniformly from $[0.75, 1.25]$, effectively stretching or compressing the audio in time and shifting its frequency content.
Second, multi-source mixing combined up to 3 noise sources at signal-to-noise ratios drawn uniformly from $[-5, 5]$~dB with probability 0.7, simulating complex multi-source acoustic environments.
Third, we augmented the secondary path impulse response used during training in three ways: (i) gain perturbation scaled the impulse response amplitude by up to $\pm 5$~dB, (ii) pairwise interpolation formed convex combinations of two secondary path impulse responses randomly sampled from the training pool, and (iii) sample shift displaced the impulse response by up to $\pm 1$ sample.
Combined, these augmentations improved noise reduction from 9.7~dB (no augmentation) to 11.2~dB (Extended Data Table~\ref{tab:aug_ablation}).

\subsection*{Hybrid partitioned convolution on DSP}

We implemented a hybrid partitioned convolution~\cite{wefers2015partitioned} on the DSP to enable efficient convolution of FIR filters without sacrificing latency. This approach first split the filter into a shorter direct segment and longer partitioned segments. We defined a fundamental block size of $B$ samples. The direct segment consisted of the first $2B$ samples of the filter and were convolved directly in the time domain. 

The remaining taps were uniformly partitioned into blocks of length $B$ and processed in the frequency domain using overlap-add with $2B$-point FFTs. To avoid stalling the real-time audio callback, the FFT-based convolution was executed in a lower priority Bela auxiliary task, while the audio thread continued streaming. During operation, we buffered the input and, once $B$ samples were accumulated, the background task computed the FFT, updated a frequency delay line, multiplied by the precomputed spectra of the partitions, and performed an inverse FFT. After the inverse FFT, we performed overlap-add in the time domain to obtain the next $B$-sample partitioned output block. The audio thread summed the time-domain output with the most recently completed partitioned block to produce the final convolution result.

In our implementation, we set the block size $B=128$. This meant we partitioned the FIR filters into 16 blocks, 2 of which were used as the direct segment to be convolved in the time domain and 14 used as the partitioned segments for frequency domain convolution.

\subsection*{Data collection and evaluation}

\header{Noise samples.} Throughout all of our experiments and evaluations, we played random samples from 11 different types of noise. These noises included car, bus, washer, heating ventilation and air conditioning (HVAC), fan, airplane cabin, mall, cafe, city, vacuum, and rain. Bus, mall, cafe, and city noises were obtained from the TAU Urban Acoustic Scenes dataset~\cite{TAUdataset}, while car noises were taken from the Vehicle Interior Sound Dataset~\cite{VISDdataset}. Rain noises were obtained from myNoise~\cite{myNoise}, and fan noises were sourced from the MIMII dataset~\cite{MIMII}. Washer and vacuum noises were obtained from the MS-SNSD dataset~\cite{MSSNSD}, and the airplane cabin noises were recorded by the authors. For each trial, every noise sample played was unique, ensuring there were no overlaps in the data.

For speech and music enhancement evaluation we used speech samples taken randomly from LibriSpeech \cite{librispeech}, and music samples from Pixabay. 

\header{Mannequin dataset.} We collected the mannequin dataset across five different rooms. \blue{In each room the mannequin was placed in two different positions, and for each position, it was rotated across the full azimuth range, stopping at 36 evenly spaced points. We placed up to 3 noise sources at random locations throughout the room, with each noise source varying in distance and height relative to the mannequin head. For each head rotation iteration we played 11 different types of noises, each as a \qty{10}{\second} clip.} The glasses were placed on the mannequin to record the reference signals, while in-ear microphones provided the ground-truth measurement of the acoustic signal at the ear. In total, we collected \qty{24}{\hour} of recordings from the mannequin head across the 5 rooms. Additional \qty{8}{\hour} of data was collected during revision for the controlled evaluations of colored noise types, alternative source loudspeakers, and anechoic conditions. These recordings were used exclusively for evaluation.

When evaluating the noise reduction performance on the mannequin head (Fig.~\ref{fig:mannequin}), we used data from three rooms for training and tested on the data from the remaining two rooms. We performed the evaluation of the mannequin head data offline.
A \qty{2}{\second} context window of recorded audio was used as the neural network input to produce the ANC filter $w_m[n]$ per channel.
To account for the inference and communication latency of an online setting, these filters were applied to the next \qty{0.5}{\second} chunk of recorded audio, delayed by \qty{200}{\milli\second} from the end of the context window.
The context window was then advanced by a hop length of \qty{0.5}{\second} to obtain the next data sample.
Specifically, we convolved the frame microphone signals $x_m[n]$ with the corresponding ANC filter $w_m[n]$ per channel and the estimated secondary path $\hat{s}[n]$ to obtain the estimated anti-noise signal reaching the ear, $\hat{d}[n]$, as defined in equation~\eqref{eq:d_hat_n}.
To evaluate the noise reduction within our operating frequency band, we applied a 4th-order Butterworth bandpass filter (\qtyrange{100}{1000}{\hertz}) to both the original ear signal $d[n]$ and the residual error signal $e[n]$ as defined in equation~\eqref{eq:e_n_vec}, and computed the noise reduction as:
\begin{equation}
    \text{Noise reduction} = 10 \log_{10} \frac{\sum_{n=0}^{N-1} \tilde{d}^2[n]}{\sum_{n=0}^{N-1} \tilde{e}^2[n]},
\end{equation}
where $\tilde{d}[n]$ and $\tilde{e}[n]$ denoted the bandpass-filtered versions of $d[n]$ and $e[n]$, respectively, and $N$ was the total number of samples in the \qty{0.5}{\second} evaluation chunk.

\header{Form factor evaluation.} We generated the synthetic dataset of the different devices using PyRoomAcoustics \cite{pyroomacoustics} simulator. Random rooms were created with lengths and widths ranging between 4 and 10 meters, and heights ranging from 2.5 to 4 meters. Wall absorption coefficients ranged from 0.1 to 0.9. A microphone array was generated for each of the devices, following the layout of Fig.~\ref{fig:mannequin}h. For each sample, a random room was generated, the microphone array was placed randomly within the room, and the TX locations were placed pseudo-randomly such that all 360 degrees of direction-of-arrival were covered. In total we generated over 90 hours of synthetic data. Two thirds of the data was used for training and one third was used for evaluation.

We developed real-world prototypes for the three form factors using MEMS microphones and speakers connected to our Bela board, similar to how we prototyped our glasses. The microphones were placed in different configurations (Fig.~\ref{fig:mannequin}h) and the speaker was placed at the temple. We placed our prototypes on the rotating mannequin head, and for each prototype collected 1 hour of training data for each of the devices in one room. We varied the transmitter and head locations across the samples. Evaluation was performed across 20 minutes for each of the devices.

\header{User evaluation.} We collected data from 16 participants organized into three groups across 12 environments.
Each participant's secondary path was estimated before data collection.
For each participant in each environment, the participant sat or stood at 3 randomly chosen locations.
\blue{At each location, the noise source was placed at 2 different random positions, with each position chosen with a random azimuth, elevation, and distance from the user.} 2 noise audio clips of \qty{10}{\second} each were played, while participants wore our glasses prototype together with the in-ear microphones for ground-truth acquisition.

\blue{For the glasses placement robustness experiment, we further tested our system's cancellation on 6 users while they performed different movements to displace the glasses. A calibration was performed before the evaluation, and the users were asked to look down, shake their heads, remove and re-wear the glasses, and pedal on an exercise bike. Throughout each movement we played ambient noise from the same noise categories from our speaker. In total, we collected \qty{2}{\hour} of data for the main study and \qty{1.5}{\hour} of data for the additional glasses-displacement study, used exclusively for evaluation.}

Noise reduction performance was evaluated using cross-group validation as described above.
During testing, the in-ear microphone was used only for ground-truth measurement and not by the deployed ANC system.
Noise reduction was computed as the power reduction between the primary noise $d[n]$ and the residual error signal $e[n]$, using the same formulation as the mannequin evaluation above.

For subjective evaluation, each of the 11 participants completed 10 trials, each using a randomly selected speech (7 trials) or music (3 trials) clip. In each trial, environmental noise was reproduced through an external loudspeaker, while the desired audio content was played through the glasses' open-ear speakers (or AirPods in the AirPods 4 condition). Each trial was structured in two phases: participants first listened to the environmental noise alone to establish a baseline impression, then heard the desired audio played back under one of three conditions: no ANC, our glasses-based ANC, or AirPods 4. \blue{To ensure that the presence of AirPods did not alter the perception of the glasses’ ANC performance, AirPods were only worn during the AirPods condition and not during the glasses conditions. The order of the three conditions was randomized across trials to minimize listener bias.} When ANC was enabled, the anti-noise signal was combined with the desired audio at the glasses speaker. After each condition, participants rated background-noise intrusiveness on a 5-point scale in response to the question: \textit{``How intrusive or noticeable were the background noises?'' (1 = most intrusive, 5 = least intrusive)}, and perceptual clarity in response to: \textit{``What was the overall listening experience and clarity of the sound?'' (1 = least clear, 5 = most clear)}.

\blue{The objective evaluation of the speech and music enhancement was conducted on the same speech and music files played for the user study. To obtain a noise-free reference, we reproduced all user-study speech and music segments through the glasses on a mannequin head in a quiet environment without background noise. This measurement does not capture per-user variation in fit and ear geometry, but provides a clean-speech reference against which the noisy and noise-reduced conditions can be interpreted. }


\subsection*{Data availability} 
All data necessary for interpreting the manuscript have been included. The datasets used in the current study are not publicly available but may be available from the corresponding authors on reasonable request and with permission of Carnegie Mellon University.

\subsection*{Code availability} 
The code used to develop our system will be made available to the public prior to publication.

\section*{Acknowledgments}
We acknowledge support from the NSF (2106921, 1942902, 2111751, 2433903), ONR, Qualcomm, and CyLab-Enterprise. Any opinions, findings, and conclusions or recommendations expressed in this material are those of the author(s) and do not necessarily reflect the views of the above.

We thank the participants at Carnegie Mellon University for their willingness to participate in the study. We thank Ashee Bansaal, Alexandra Yin, Jiangyifei Zhu, Leah Zhang, Seungjoo Lee, Siqi Zhang, Veronica Muriga, Yawen Liu, Zhikai Qin, and Tuochao Chen for their critical and important feedback on the manuscript. We thank Tarun Pruthi and Xiaoran Fan for their suggestions when conceptualizing the study.

\section*{Author contributions}
K.Y., F.L., T.X., J.C. and S.K. conceptualized the study; K.Y. and F.L. designed the system, performed the dataset collection and human subjects study, conducted the experiments, and performed the analysis under technical supervision by S.K. and J.C.; K.Y., F.L., Y.S., C.S. and S.B. developed the system prototype; K.Y., F.L., T.X, J.C. and S.K. wrote the manuscript. 

\section*{Competing interests}
The authors declare the following competing interests: J.C. is a co-founder of Wavely Diagnostics, Inc. The remaining authors declare no competing interests.

\section*{Supplementary Materials}
Supplementary Video 1: Ambient noise reduction while playing music, demonstrated on a street and inside a car.\\
Supplementary Video 2: Ambient noise reduction while playing speech, demonstrated on a bus and in a kitchen.\\
Supplementary Video 3: Noise reduction under user movement and a mid-recording change of noise source.\\
Supplementary Video 4: Noise reduction comparison between in-ear and outside-ear placement, and the user-specific calibration procedure.\\

\bibliography{references}
\bibliographystyle{naturemag}

\renewcommand{\figurename}{Extended Data Figure}
\setcounter{figure}{0}
\setcounter{table}{0}
\renewcommand{\tablename}{Extended Data Table}

\clearpage
\newpage
\begin{figure}[h]
    \centering
        \caption{\blue{Power spectral density (PSD) of noise at the ear with ANC off and on across different real-world and colored noise types.}}
    \includegraphics[width=0.97\linewidth]{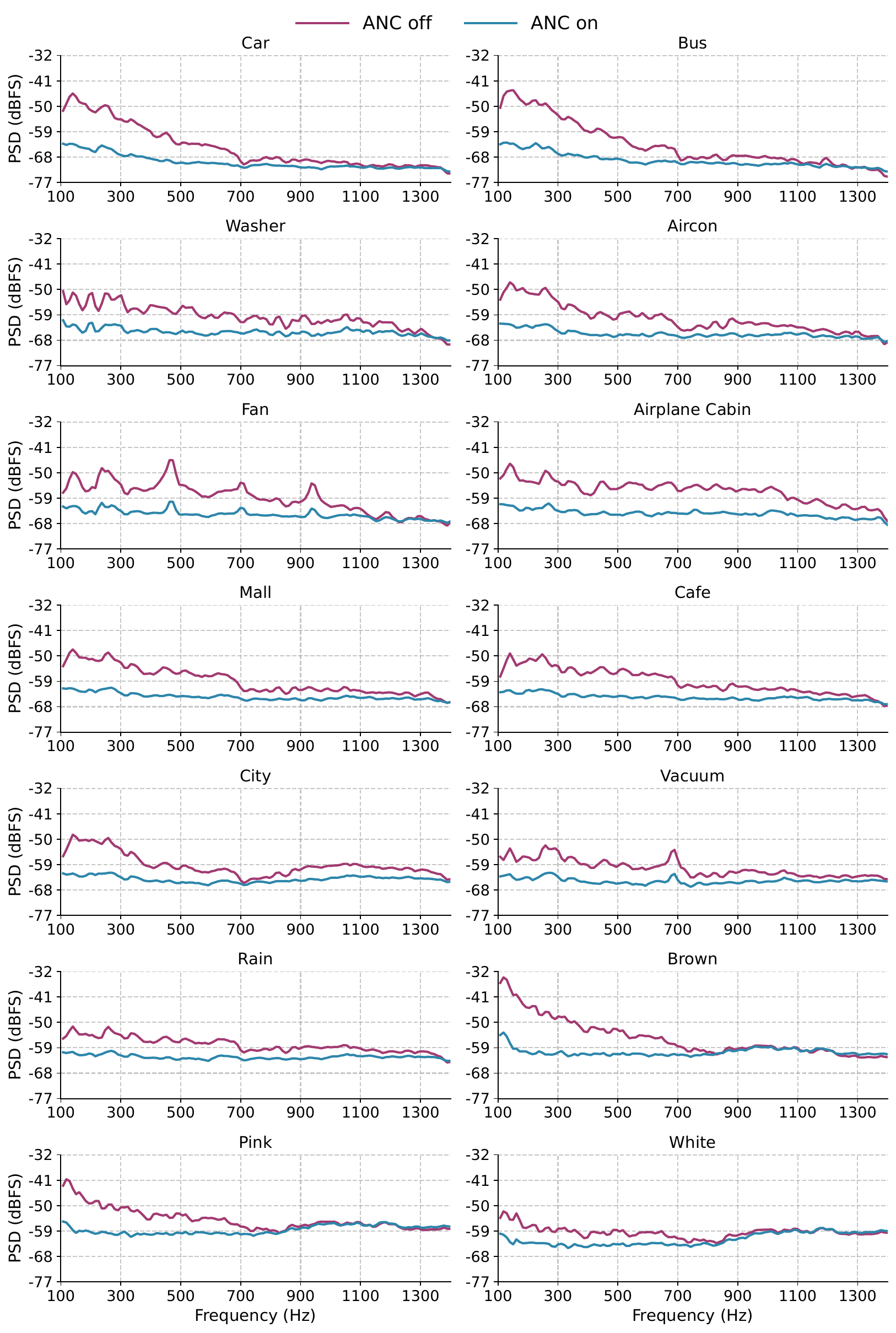}
    \label{fig:noise_type_psd}
\end{figure}

\clearpage
\newpage
\begin{figure}[h]
    \centering
        \caption{\textbf{Effect of microphone count and placement on noise reduction.} \textbf{a,} Best-performing microphone configurations for each channel count. Orange circles with 1--6 indicate the six microphone locations. \textbf{b,} Noise reduction across directions of arrival for different microphone counts. \textbf{c,} Comparison of two-microphone configurations, showing that spatially distributed placements (Mics 1\&2, 1\&3) outperform clustered placements (Mics 3\&4) by providing richer spatial information. A smaller radius indicates greater noise reduction.}
    \includegraphics[width=0.9\linewidth]{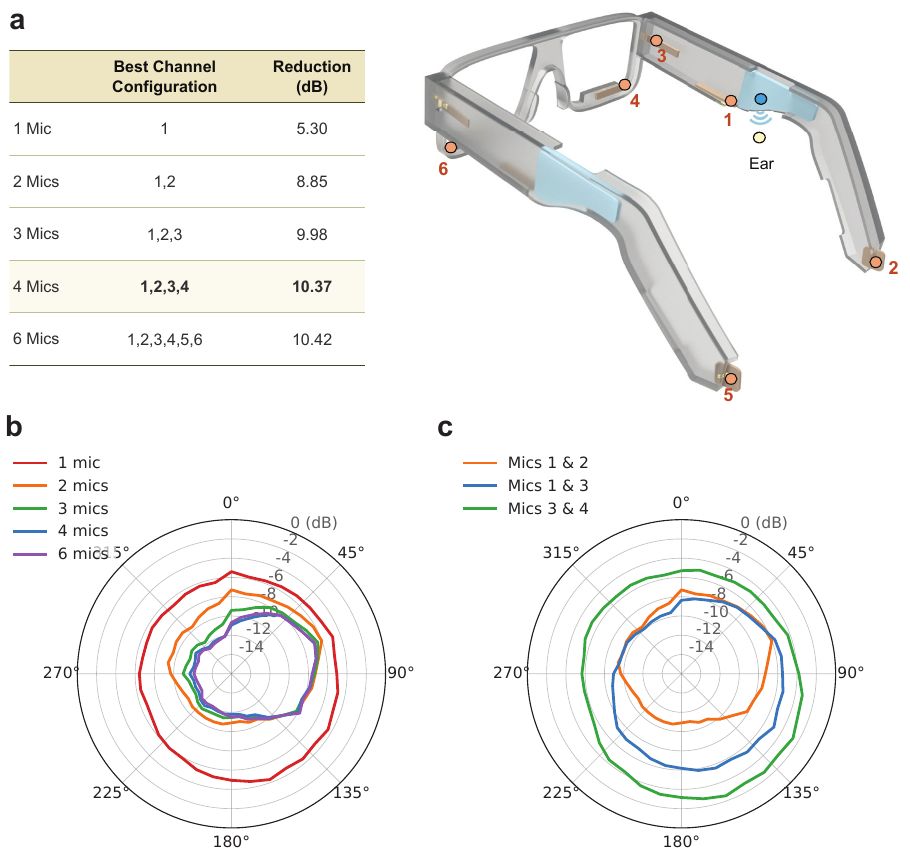}

    \label{fig:mic_config}
\end{figure}

\clearpage
\newpage
\begin{figure}[h]
    \centering
    \caption{\blue{\textbf{a.} Frequency response of the different tested loudspeakers. \textbf{b.} Performance of our system with the different loudspeakers playing the primary noise. Our system maintains meaningful noise reduction across all three different speakers.}}
    \includegraphics[width=\linewidth]{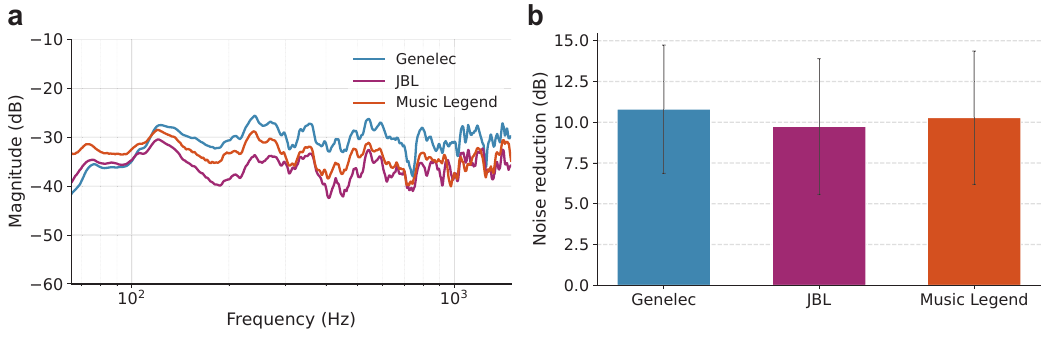}
    \label{fig:source_speaker}
\end{figure}

\clearpage
\newpage
\begin{figure}[h]
    \centering
    \caption{\blue{\textbf{Noise reduction under glasses displacement.} Noise reduction performance of our system under different user movements to displace the glasses. Calibration was only done once at the beginning and not between movements. Even with the user deliberately displacing the glasses, our system's mean noise reduction stays within 0.6~dB of the calibrated performance.}}
    \includegraphics[width=0.65\linewidth]{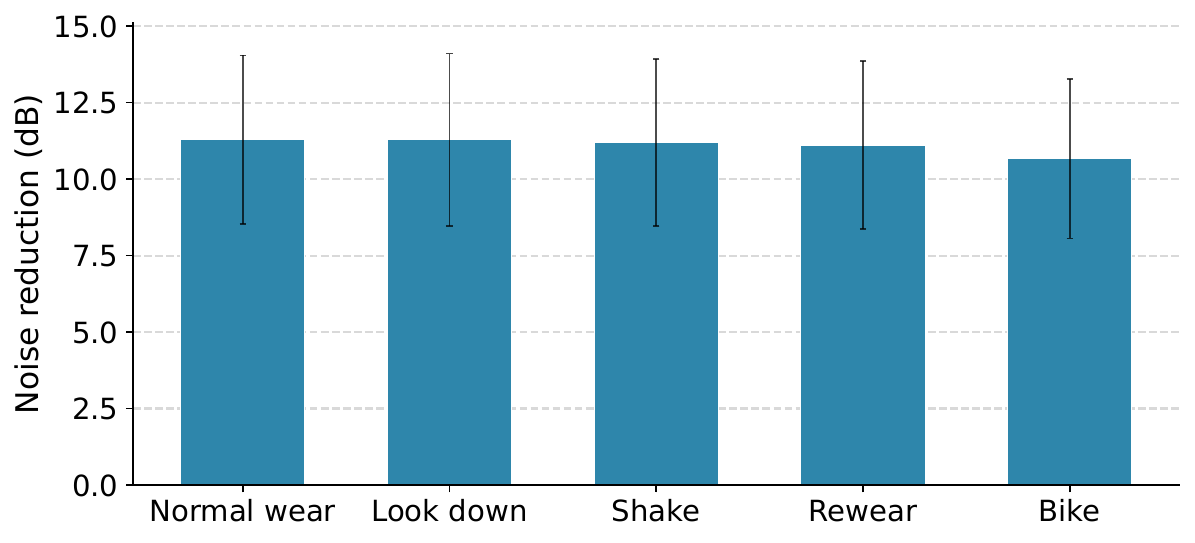}
    \label{fig:glasses_displacement}
\end{figure}

\clearpage
\newpage
\begin{figure}[h]
    \centering
    \caption{\blue{\textbf{Effect of background-noise level during secondary path calibration on noise reduction performance.} Noise reduction achieved when the secondary path is calibrated under different signal-to-noise ratios between the exponential sine sweep stimulus and background noise. Performance is consistent from 30 to 0~dB SNR, and degrades by less than 1.7~dB at -20 dB SNR, showing the calibration does not require a quiet environment. Error bars show the standard deviation}}
    \includegraphics[width=0.65\linewidth]{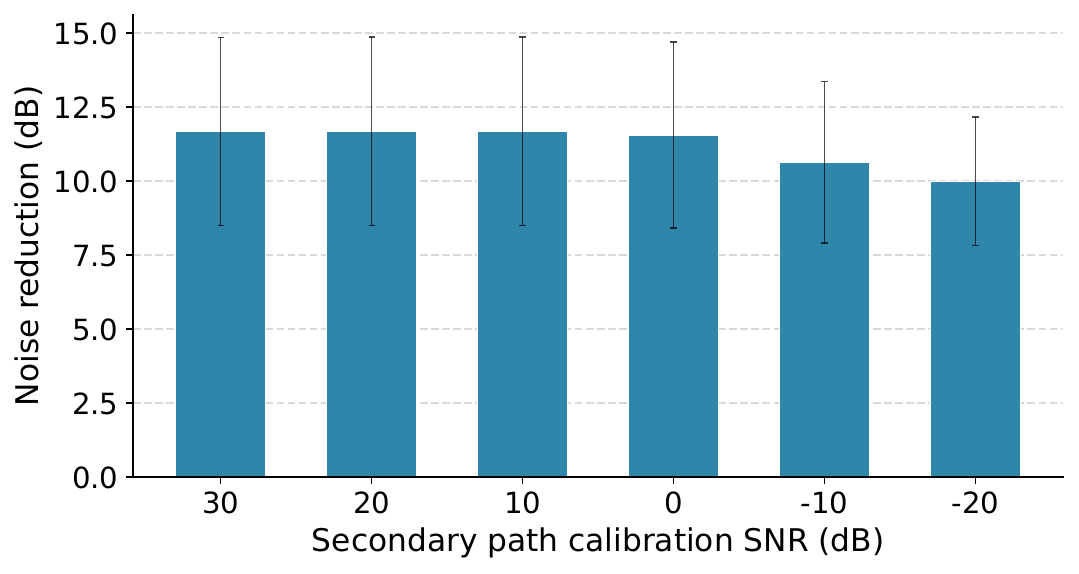}
    \label{fig:calibration_snr}
\end{figure}



\clearpage
\newpage
\begin{figure}[h]
    \centering
    \caption{Effect of DSP computation latency on noise reduction performance. Latency is varied from 45~\unit{\micro\second} ($1/f_s$) to 726~\unit{\micro\second} ($16/f_s$), where $f_s$~= 22050~Hz is the audio sampling rate, showing that reduction degrades as processing delay increases.}
    \includegraphics[width=0.65\linewidth]{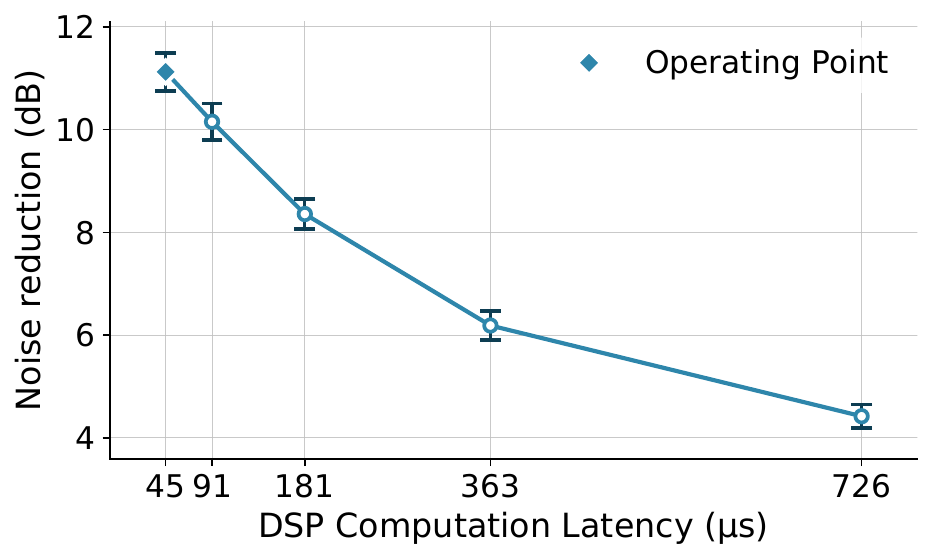}
    \label{fig:delay_noise_reduction}
\end{figure}

\clearpage
\newpage
\begin{figure}[h]
    \centering
    \caption{\textbf{Custom microphone breakout board design.} \textbf{a,} Off-the-shelf Sparkfun breakout board for the SPH8878LR5H-1 MEMS microphone. \textbf{b,} Custom breakout board that uses the same circuit schematic but with a redesigned PCB layout, allowing the microphone to fit within the narrow edges of the 3D-printed glasses frame.}
    \includegraphics[width=0.65\linewidth]{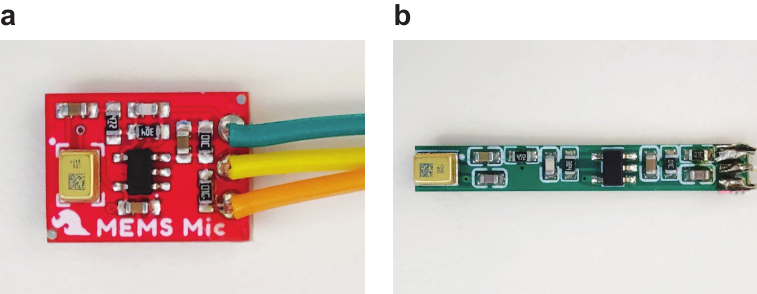}
    \label{fig:mic_pcb}
\end{figure}

\newpage
\begin{table}[H]
\centering
\caption{\textbf{Acoustic characteristics of the evaluation environments.} The table summarizes the physical dimensions and the reverberation time (RT60) for each room used in our real-world evaluation.}
\label{tab:room_stats}
\begin{tabular}{@{}lccc@{}}
\toprule
\textbf{Room ID} & \textbf{Environment Type} & \textbf{Dimensions ($L \times W \times H$ m)} & \textbf{RT60 (s)} \\ \midrule
R1               & Conference Room           & $8.7 \times 7.0 \times 2.9$                   & 0.52                   \\
R2               & Office                    & $5.0 \times 3.5 \times 3.6$                   & 0.50                   \\
R3               & Lobby                     & $10.0 \times 8.0 \times 12.0$                 & 0.90                   \\
R4               & Kitchen                    & $7.0 \times 5.6 \times 3.6$                   & 0.53                   \\
R5               & Conference Room             & $4.8 \times 3.6 \times 2.6$                                            & 0.52                   \\
R6               & Lecture Hall                 & $15.0 \times 12.0 \times 3.3$                 & 0.48                   \\
R7               & Conference Room                & $7.0 \times 4.8 \times 2.6$                   & 0.63                   \\ \bottomrule
\end{tabular}
\end{table}

\clearpage
\newpage
\begin{table}[h]
\centering
\caption{Ablation study on model architecture and size. LSTM$_h$: hidden dimension of the LSTM layer; SE: Squeeze-and-Excitation blocks; Ref.: reference microphone input; MACs: multiply--accumulate operations per second; NR: overall noise reduction in the target 100--1000\,Hz frequency band.}
\label{tab:ablation}
\setlength{\tabcolsep}{4pt}
\begin{tabular}{@{}lccccccc@{}}
\toprule
Configuration & Encoder Channels & LSTM$_h$ & SE & Ref. & Params & MACs & NR (dB) \\
\midrule
\multicolumn{8}{@{}l}{\textit{Component ablation}} \\[2pt]
\quad w/o LSTM \& Ref. & [32,64,64,128,128,256,256] & --  & \checkmark &         & 3.3M & 5.3G & 9.0 \\
\quad w/o Ref. input    & [32,64,64,128,128,256,256] & 128  & \checkmark &         & 4.2M & 5.3G & 9.1 \\
\quad w/o LSTM & [32,64,64,128,128,256,256] & --  & \checkmark & \checkmark        & 3.3M & 5.3G  & 10.3 \\
\quad w/o SE blocks     & [32,64,64,128,128,256,256] & 128  &         & \checkmark & 4.1M & 5.3G & 10.3 \\
\quad \textbf{Base}    & [32,64,64,128,128,256,256] & 128 & \checkmark & \checkmark & \textbf{4.2M} &  \textbf{5.3G} & \textbf{10.4} \\
\midrule
\multicolumn{8}{@{}l}{\textit{Model size scaling}} \\[2pt]
\quad Small   & [16,32,32,64,64,128,128]   &  64 & \checkmark & \checkmark & 1.1M &  1.4G &  9.7 \\
\quad Medium  & [24,48,48,96,96,192,192]     & 96 & \checkmark & \checkmark & 2.3M &  3.1G & 10.2 \\
\quad \textbf{Base}    & [32,64,64,128,128,256,256] & 128 & \checkmark & \checkmark & \textbf{4.2M} &  \textbf{5.3G} & \textbf{10.4} \\
\quad Large   & [48,96,96,192,192,384,384] & 192 & \checkmark & \checkmark & 9.4M & 11.8G & 10.6 \\
\bottomrule
\end{tabular}
\end{table}

\clearpage
\newpage
\begin{table}[h]
\centering
\caption{Performance comparison when training the neural network on different datasets}
\begin{tabular}{l c c}
\hline
\multirow{2}{*}{\textbf{Training strategy}} & \multicolumn{2}{c}{\textbf{Noise reduction (dB)}} \\
\cline{2-3}
 & With calibration & Without calibration \\
\hline
Mannequin data only & 10.4 & 7.6\\
Human data only  &  10.1 & 9.1 \\
Pre-train with mannequin data, fine-tune with human data & \textbf{11.2} & \textbf{9.6} \\
\hline
\end{tabular}
\label{tab:my_label}
\end{table}

\clearpage
\newpage
\begin{table}[h]
\centering
\caption{Ablation study on data augmentation strategies (SP: Secondary path).}
\label{tab:aug_ablation}
\setlength{\tabcolsep}{5pt}
\begin{tabular}{@{}lcccccc@{}}
\toprule
Configuration & Speed & Mixing & SP Gain & SP Interp. & SP Shift & Noise Reduction (dB) \\
\midrule
\quad No aug.       & & & & & & 9.7 \\
\quad SP aug.\ only         & & & \checkmark & \checkmark & \checkmark & 10.1 \\
\quad Audio aug.\ only      & \checkmark & \checkmark & & & & 10.3 \\
\quad All aug.       & \checkmark & \checkmark & \checkmark & \checkmark & \checkmark & \textbf{11.2} \\
\midrule
\multicolumn{7}{@{}l}{\textit{Audio augmentation ablation}} \\[2pt]
\quad Mixing only              & & \checkmark & \checkmark & \checkmark & \checkmark & 10.3 \\
\quad Speed only            & \checkmark & & \checkmark & \checkmark & \checkmark & 11.1 \\
\quad Speed + Mixing           & \checkmark & \checkmark & \checkmark & \checkmark & \checkmark & \textbf{11.2} \\
\midrule
\multicolumn{7}{@{}l}{\textit{Secondary path (SP) augmentation ablation}} \\[2pt]
\quad SP Shift only         & \checkmark & \checkmark & & & \checkmark & 10.4 \\
\quad SP Interp. only        & \checkmark & \checkmark & & \checkmark & & 10.7 \\
\quad SP Gain only          & \checkmark & \checkmark & \checkmark & & & 11.1 \\
\quad All SP aug.     & \checkmark & \checkmark & \checkmark & \checkmark & \checkmark & \textbf{11.2} \\
\bottomrule
\end{tabular}
\end{table}

\clearpage
\newpage
\begin{table}[h]
\centering
\caption{\blue{Noise reduction as a function of the number of noise sources in the environment, and the number of microphone channels utilized, compared to the Wiener filter performance. The Wiener filter provides an oracle upper bound to our system as it uses the in-ear microphone for noise reduction. When evaluating the number of noise sources, 4 microphone channels were used. When evaluating the number of microphone channels, 1 noise source was used. 
Values are mean $\pm$ standard deviation (dB).}}
\label{tab:wiener}
\setlength{\tabcolsep}{8pt}
\begin{tabular}{@{}lcc@{}}
\toprule
 & \textbf{Ours} & \textbf{Wiener Filter with In-ear Mic.} \\
\midrule
\multicolumn{3}{@{}l}{\textit{Number of noise sources}} \\[2pt]
\quad 1 source  & $10.4 \pm 3.5$ & $14.7 \pm 4.7$ \\
\quad 2 sources & $9.7 \pm 3.3$ & $13.3 \pm 4.2$ \\
\quad 3 sources & $9.4 \pm 3.2$ & $12.6 \pm 4.0$ \\
\midrule
\multicolumn{3}{@{}l}{\textit{Number of microphone channels}} \\[2pt]
\quad 1 mic & $5.3 \pm 2.8$ & $6.6 \pm 3.4$ \\
\quad 2 mic & $8.9 \pm 3.5$ & $11.7 \pm 4.5$ \\
\quad 3 mic & $10.0 \pm 3.5$ & $14.1 \pm 4.6$ \\
\quad 4 mic & $10.4 \pm 3.5$ & $14.7 \pm 4.7$ \\
\quad 6 mic & $10.4 \pm 3.6$ & $15.5 \pm 5.0$ \\
\bottomrule
\end{tabular}
\end{table}


\end{document}